\documentclass[11pt]{article}
\usepackage{slashed}
\usepackage{amsfonts}
\usepackage{amsmath}
\usepackage{amssymb}
\usepackage{hyperref}
\usepackage{epsfig}

\usepackage{pgf,tikz}
\usetikzlibrary{calc}

\usetikzlibrary{arrows}
\usetikzlibrary{patterns}
\usepackage{mathrsfs}
\catcode`\@11\relax
\newif\ify@autoscale \y@autoscaletrue \def\Yautoscale#1{\ifnum #1=0
  \y@autoscalefalse\else\y@autoscaletrue\fi}
\newdimen\y@b@xdim
\newdimen\y@boxdim \y@boxdim=13pt
\def\Yboxdim#1{\y@autoscalefalse\y@boxdim=#1}
\newdimen\y@linethick    \y@linethick=.3pt
\def\Ylinethick#1{\y@linethick=#1}
\newskip\y@interspace \y@interspace=0ex plus 0.3ex
\def\Yinterspace#1{\y@interspace=#1}
\newif\ify@vcenter   \y@vcenterfalse
\def\Yvcentermath#1{\ifnum #1=0 \y@vcenterfalse\else\y@vcentertrue\fi}
\newif\ify@stdtext   \y@stdtextfalse
\def\Ystdtext#1{\ifnum #1=0 \y@stdtextfalse\else\y@stdtexttrue\fi}
\newif\ify@enable@skew   \y@enable@skewfalse
\expandafter\ifx\csname enableskew\endcsname\relax
 \y@enable@skewfalse \else \y@enable@skewtrue\fi
\def\y@vr{\vrule height0.8\y@b@xdim width\y@linethick depth 0.2\y@b@xdim}
\def\y@emptybox{\y@vr\hbox to \y@b@xdim{\hfil}}
\ify@enable@skew
 \def\y@abcbox#1{\if :#1\else
   \y@vr\hbox to \y@b@xdim{\hfil#1\hfil}\fi}
 \def\y@mathabcbox#1{\if :#1\else
   \y@vr\hbox to \y@b@xdim{\hfil$#1$\hfil}\fi}
\else
 \def\y@abcbox#1{\y@vr\hbox to \y@b@xdim{\hfil#1\hfil}}
 \def\y@mathabcbox#1{\y@vr\hbox to \y@b@xdim{\hfil$#1$\hfil}}
\fi
\def\y@setdim{%
  \ify@autoscale%
   \ifvoid1\else\typeout{Package youngtab: box1 not free! Expect an
     error!}\fi%
   \setbox1=\hbox{A}\y@b@xdim=1.6\ht1 \setbox1=\hbox{}\box1%
  \else\y@b@xdim=\y@boxdim \advance\y@b@xdim by -2\y@linethick
  \fi}
\newcount\y@counter
\newif\ify@islastarg
\def\y@lastargtest#1,#2 {\if\space #2 \y@islastargtrue
  \else\y@islastargfalse\fi}
\def\y@emptyboxes#1{\y@counter=#1\loop\ifnum\y@counter>0
  \advance\y@counter by -1 \y@emptybox\repeat}
\def\y@nelineemptyboxes#1{%
  \vbox{%
    \hrule height\y@linethick%
    \hbox{\y@emptyboxes{#1}\y@vr}
    \hrule height\y@linethick}\vskip-\y@linethick}
\def\yng(#1){%
  \y@setdim%
  \hskip\y@interspace%
  \ifmmode\ify@vcenter\vcenter\fi\fi{%
  \y@lastargtest#1,
  \vbox{\offinterlineskip
    \ify@islastarg
     \y@nelineemptyboxes{#1}
    \else
     \y@ungempty(#1)
    \fi}}\hskip\y@interspace}
\def\y@ungempty(#1,#2){%
  \y@nelineemptyboxes{#1}
  \y@lastargtest#2,
  \ify@islastarg
   \y@nelineemptyboxes{#2}
  \else
   \y@ungempty(#2)
  \fi}
\def\y@nelettertest#1#2. {\if\space #2 \y@islastargtrue
  \else\y@islastargfalse\fi}
\def\y@abcboxes#1#2.{%
  \ify@stdtext\y@abcbox#1\else\y@mathabcbox#1\fi%
  \y@nelettertest #2.
  \ify@islastarg\unskip%
   \ify@stdtext\y@abcbox{#2}\else\y@mathabcbox{#2}\fi%
  \else\y@abcboxes#2.\fi}
 \newdimen\y@full@b@xdim
 \newcount\y@m@veright@cnt
\ify@enable@skew
 \def\y@get@m@veright@cnt#1#2.{%
   \if :#1 \advance\y@m@veright@cnt by 1\y@get@m@veright@cnt#2.\fi}
 \let\y@setdim@=\y@setdim
 \def\y@setdim{%
   \y@setdim@ \y@full@b@xdim=\y@b@xdim
   \advance\y@full@b@xdim by 1\y@linethick}
 \def\y@m@veright@ifskew#1{
   \y@m@veright@cnt=0 \y@get@m@veright@cnt#1.
   \moveright \y@m@veright@cnt\y@full@b@xdim}
\else
 \def\y@m@veright@ifskew#1{}
\fi
\def\y@nelineabcboxes#1{%
  \y@nelettertest #1.
  \ify@islastarg
   \y@m@veright@ifskew{#1}
    \vbox{
      \hrule height\y@linethick%
      \hbox{\ify@stdtext\y@abcbox#1\else\y@mathabcbox#1\fi\y@vr}
      \hrule height\y@linethick}\vskip-\y@linethick
  \else
   \y@m@veright@ifskew{#1}
    \vbox{
      \hrule height\y@linethick%
      \hbox{\y@abcboxes #1.\y@vr}%
      \hrule height\y@linethick}\vskip-\y@linethick
  \fi}
\def\young(#1){%
  \y@setdim%
  \hskip\y@interspace%
  \y@lastargtest#1,
  \ifmmode\ify@vcenter\vcenter\fi\fi{%
  \vbox{\offinterlineskip
    \ify@islastarg\y@nelineabcboxes{#1}%
    \else\y@ungabc(#1)%
    \fi}}\hskip\y@interspace}
\def\y@ungabc(#1,#2){%
  \y@nelineabcboxes{#1}%
  \y@lastargtest#2,
  \ify@islastarg\y@nelineabcboxes{#2}%
  \else\y@ungabc(#2)%
  \fi}
\catcode`\@12\relax
 
\usepackage{geometry} 
\usepackage{graphicx}
\usepackage{tabularx}
\usepackage{typearea}
\usepackage{wrapfig}
\usepackage{amsthm}

\usepackage{accents}
\usepackage[ansinew]{inputenc}
\usepackage[arrow, matrix, curve]{xy}
\usepackage{enumerate}
\usepackage{bbm}
\usepackage{mathtools}
\usepackage{cite}
\usepackage{tikz}
\usepackage{etex}
\usepackage{tikz-qtree}
\usetikzlibrary{trees}
\usetikzlibrary{calc}

\numberwithin{equation}{section}

\newtheorem{thm}{Theorem}[section]

\newtheorem{lem}[thm]{Lemma}

\newtheorem{rem}[thm]{Remark}
\newtheorem{exx}[thm]{Example}

\theoremstyle{theorem}
\newtheorem{Theorem}{Theorem}[section]
\newtheorem{Proposition}[Theorem]{Proposition}
\newtheorem{Lemma}[Theorem]{Lemma}
\newtheorem{Corollary}[Theorem]{Corollary}
\newtheorem{Definition}[Theorem]{Definition}

\theoremstyle{remark}
\newtheorem{Remark}[Theorem]{Remark}
\newtheorem{Example}[Theorem]{Example}
\newcommand{\bco}{\begin{Corollary}}
\newcommand{\eco}{\end{Corollary}}

\newcommand{\unit}{{\rm 1\hspace*{-0.4ex}%
\rule{0.1ex}{1.52ex}\hspace*{0.2ex}}}

\newcommand{\C}{{\mathbb C}}

\newcommand{\ulj}{\underline{j}}
\newcommand{\ulk}{\underline{k}}
\newcommand{\ulK}{\underline{K}}
\newcommand{\ulJ}{\underline{J}}

\newcommand{\cdt}{\!\cdot \!}
\newcommand{\sz}{{\textsf{z}}}

\newcommand{\sx}{{\textsf{x}}}

\newcommand{\sy}{{\textsf{y}}}

                                                                                                                                                                                                                                                                                                                                                                       \newcommand{\ulm}{\underline{m}}

\newcommand{\ulu}{\underline{u}}

\newcommand{\chibig}{\scalebox{1.2}{$\chi$}}
\newcommand{\boplus}{\scalebox{1.2}{$\oplus$}}
\def\boldsig{\mbox{\boldmath{$\sigma$}}}
\def\bVee{\scalebox{1.3}{\boldmath{$\vee$}}}

\def\srbVee{\scalebox{1.2}{\boldmath{$\vee$}}}

\def\bWee{\bVee\hskip-1.5ex\raisebox{-1.4ex}{$\srbVee$}}

\def\bcdot{\mbox{\scalebox{1.4}{$\cdot$}}}
\def\fh{\textstyle{\frac12}}
\def\fth{\textstyle{\frac32}}
\def\boldsig{\mbox{\boldmath{$\sigma$}}}

\def\hj{\textstyle{\frac 12}j}
\def\one{\hbox{{1}\kern-.25em\hbox{l}}}
\def\hj{\textstyle{\frac 12}j}

\def\bJ{\mathbb{J}}
\def\bfJ{\mathbf J}
\def\bP{\mathbb{P}}

\def\hj{\textstyle{\frac 12}j}

\def\oh{\textstyle{\frac 12}}

\def\boldsig{\mbox{\boldmath{$\sigma$}}}

\newcommand\nj[9]
{\left\{\begin{array}{ccc}#1&#2&#3\\#4&#5&#6\\#7&#8&#9\end{array}\right\}}

\newcommand\scb[2]{\scalebox{#1}{#2}}
\def\dtnj{\nj{\cdot}{\cdot}{\cdot}{\cdot}{\cdot}{\cdot}{\cdot}{\cdot}{\cdot}}

\def\sp{\!+\!}
\def\sm{\!-\!}

\def\bx{\mbox{}\hfill$\Box$}
\def\qed{\bx}
\def\qeb{\bx}
\newcommand{\BF}[3]{(\chibig_{#1})^{#2}_{#3}}

\newcommand{\BFi}[1]{\chi_{#1}}
\newcommand{\BFCi}[1]{\chi_{#1}^{\CC}}
\newcommand{\bpr}{\begin{Proposition}}
\newcommand{\epr}{\end{Proposition}}

\newcommand{\btm}{\begin{Theorem}}
\newcommand{\etm}{\end{Theorem}}

\newcommand{\ben}{\begin{enumerate}}
\newcommand{\een}{\end{enumerate}}

\newcommand{\bit}{\begin{itemize}}
\newcommand{\eit}{\end{itemize}}

\newcommand{\bca}{\begin{cases}}
\newcommand{\eca}{\end{cases}}

\newcommand{\bre}{\begin{Remark}\rm}
\newcommand{\ere}{\end{Remark}}

\newcommand*{\bbm}{\begin{Remark}}
\newcommand*{\ebm}{\end{Remark}}

\newcommand{\ble}{\begin{Lemma}}
\newcommand{\ele}{\end{Lemma}}

\newcommand*{\bsz}{\begin{Proposition}}
\newcommand*{\esz}{\end{Proposition}}

\newcommand{\beq}{\begin{equation}}
\newcommand{\eeq}{\end{equation}}

\newcommand{\bbma}{\begin{bmatrix}}
\newcommand{\ebma}{\end{bmatrix}}

\newcommand{\bpma}{\begin{pmatrix}}
\newcommand{\epma}{\end{pmatrix}}

\newcommand*{\bbs}{\begin{Example}}
\newcommand*{\ebs}{\end{Example}}

\newcommand*{\bfg}{\begin{Corollary}}
\newcommand*{\efg}{\end{Corollary}}

\newcommand*{\bdf}{\begin{Definition}}
\newcommand*{\edf}{\end{Definition}}

\newcommand*{\bbw}{\begin{proof}}
\newcommand*{\ebw}{\end{proof}}

\newcommand*{\bpf}{\begin{proof}}
\newcommand*{\epf}{\end{proof}}

\newtagform{simple}{}{}
\newtagform{standard}{(}{)}

\allowdisplaybreaks

\newcommand{\II}{\mathbbm{1}}

\newcommand{\CC}{{\mathbb{C}}}

\newcommand{\PK}{\mathbb{P}}

\newcommand{\F}{\mathbb{F}}
\newcommand{\pr}{\mr{pr}}

\newcommand{\ti}{\tilde}

\newcommand{\SU}{{\mr{SU}}}
\newcommand{\su}{\mf{su}}

\newcommand{\al}[1]{\begin{align} #1 \end{align}}
\newcommand{\ala}[1]{\begin{align*} #1 \end{align*}}

\DeclareMathOperator{\Tr}{tr}

\newcommand{\mc}[1]{\mathcal{#1}}
\newcommand{\mf}[1]{\mathfrak{#1}}
\newcommand{\mr}[1]{\mathrm{#1}}

\newcommand{\comment}[1]{}
\newcommand{\verweis}[1]{}
\newcommand{\todo}[1]{}

\newcommand{\ket}[1]{|#1\rangle}
\newcommand{\bra}[1]{\langle#1|}
\newcommand{\braket}[2]{\langle#1|#2\rangle}

\newcommand{\vp}{\varphi}

\newcommand{\ctg}{\mr T^\ast}

\newcommand{\rref}[1]{{\rm \ref{#1}}}

\newcommand{\ol}[1]{\overline{#1}}
\newcommand{\ul}[1]{\underline{#1}}

\begin{document}

\title{\protect 
{\Large \textsc{Spin chain techniques for angular momentum quasicharacters.}   }  }

\author{P D Jarvis\footnote{School of Natural Sciences (Mathematics and Physics), University of Tasmania \texttt{(peter.jarvis@utas.edu.au)}}, and
G Rudolph${}^\ddagger$ 
\footnote{Institut  f\"{u}r Theoretische Physik, Universit\"{a}t Leipzig \texttt{(rudolph@rz.uni-leipzig.de)}}, 
}

\maketitle

\abstract

We study the ring of invariant functions over the $N$-fold Cartesian product of copies of the compact Lie group $G=\SU(2)$, modulo the action of conjugation by the diagonal subgroup, generalizing the group character ring.  For $N=1$, an orthonormal basis for the space of invariant functions is given by the irreducible characters, and the structure constants under pointwise multiplication are the coefficients of the Clebsch-Gordan series for the reduction of angular momentum tensor products ($3j$ coefficients). For $N \ge 2$\,, the structure constants under pointwise multiplication of the corresponding invariants, which we term irreducible quasicharacters, are Racah $3(2N\!-\!1)j$ recoupling coefficients, which can be decomposed as products of $9j$ coefficients (for $N=2$, they are squares thereof).
We identify the irreducible quasicharacters for $\times^N\! \SU(2)$ with traces of representations of group elements, over totally coupled angular momentum states labelled by binary coupling trees $T$ with $N$ leaves, $N\!-\!1$ internal vertices and associated intermediate edge labels. Using concrete spin chain realizations and projection techniques, we give explicit constructions for some low degree $N=2, 3$ and $4$ quasicharacters.
In the case $N=2$, related methods
are used to work out the expansions of products of generic, with elementary spin-$\textstyle{\frac 12}$, quasicharacters (equivalent to an \emph{ab initio} evaluation of certain basic $9j$ coefficients)\,. We provide an appendix which summarizes
formal properties of the quasicharacter calculus known from our previous work for both $\SU(2)$ and for compact $G$ (J Math Phys \underline{59}(8) 083505 (2018)  and 
\underline{62}(3) 033514 (2021)\,. In particular, we provide an explicit  derivation for the $N=2$ angular momentum quasicharacter product rule.


\mbox{}\\
\vfill
\pagebreak
\tableofcontents
\listoftables
\pagebreak

\section{Introduction.}
\label{sec:Introduction}

The present work is closely related to our previous papers \cite{fuerstenberg2017defining,Fuchs2018costratificationMR,jarvis2021quasicharactersMR}, which  
are part of a program that aims at developing a non-perturbative quantum theory of gauge fields in the Hamiltonian framework with special emphasis on the role of non-generic gauge orbit types. In this approach, the starting point is a finite-dimensional  lattice approximation of the classical Hamiltonian model. This approximation yields a Hamiltonian system endowed with 
the action of (the lattice counterpart of) the group of local gauge transformations. If the gauge group $G$ is non-Abelian, then this action necessarily has more than one orbit type. Correspondingly, the reduced phase space of the Hamiltonian system, obtained by symplectic reduction, is a stratified symplectic space. The corresponding quantum theory is obtained via canonical quantization. It is best described in the language of $C^*$-algebras (see \cite{qcd3, GR2}). 
To study the influence of the classical orbit type stratification at the quantum level, the concept of costratification of the quantum Hilbert space as developed by Huebsch\-mann \cite{Hue:Quantization} is used. The latter is naturally implemented within the framework 
of holomorphic quantization (a case study is given in \cite{HRS}). Within this holomorphic picture, the relations characterizing the classical 
gauge orbit strata may be implemented at the quantum level. Thereby, each element of the stratification corresponds to the zero locus of a finite subset $\{p_1, \ldots, p_r\}$ of the algebra $\mc R(G)$ of $G$-invariant representative functions on $G^N_\CC$, which will be referred to as quasicharacters.\footnote{Here, $G_\CC$ denotes the complexification of $G$.} Via this route, we were led to study the algebra $\mc R(G)$. This was done in \cite{Fuchs2018costratificationMR} and   \cite{jarvis2021quasicharactersMR} for $G = \SU(2)$ and, subsequently,  for an arbitrary compact Lie group $G$. For $G= \SU(2)$, our analysis boils down to a problem in the combinatorics of angular momentum theory. For the latter we refer to  \cite{biedenharn1981angular,biedenharn1981racah,louck2008unitary,Yutsis,wormer2006angular}. Using this theory, 
the multiplicative structure constants of $\mc R(G)$ may be expressed in terms of Clebsch-Gordan coefficients. As a consequence, we have obtained a characterization of the costrata of quantum theory in terms of systems of linear equations with real coefficients built from Wigner's $3nj$ symbols. The latter may be further expressed in terms of $9j$ symbols. For these symbols there exist efficient calculators, that is, the above coefficients can be calculated explicitly. Using the same techniques, we have been also able to reduce the eigenvalue problem for the quantum Hamiltonian to a problem in linear algebra. 

Although angular momentum theory yields an effective (computer supported) calculus, we believe that it is interesting  to 
relate the latter to the calculus of classical invariant theory. This is the aim of the present paper. The main tools used for 
accomplishing our goal come from the theory of characteristic identities, see  \cite{green1971characteristic,bracken1971vector}, \cite{OBrien1977}, \cite{gould1978tensor}, \cite{jarvis1979casimir},  \cite{gould1985characteristic,isaac2015characteristic}. 
For a given reduction scheme of the tensor product \smash{$(D_{\ul j}, H_{\ul j})$} of $N$ irreps of $\SU(2)$, this theory provides a family of operators \smash{${\mathbb P}^j_{\ul j, \ul l}$}  projecting on to the irreducible subspaces \smash{$H_{\ul j , \ul l}$ of  $H_{\ul j}$}. It turns out that 
the quasicharacters of $D_{\ul j}$ may be expressed in terms of traces of the form 
\smash{$\mathrm {Tr} \big( {\mathbb P}^j_{\ul j, \ul l'}  D_{\ul j}(\ul u)  {\mathbb P}^j_{\ul j, \ul l } \big)$}. This observation is the starting point of our calculus, which aims at expressing the quasicharacters in terms of ordinary trace invariants of group elements. Moreover, we 
are also able to analyze the multiplication laws for quasicharacters along these lines.  In the present paper, we only analyze certain 
types of products. A general analysis for quasicharacter products with distinct group elements is left to future work. For that purpose, we will  have to combine the methods used in the present paper with recoupling calculus based on different tree labelling. See Lemma \ref{lem:PtWiseT1T1Split} for some insight.

An outline of the paper is as follows. In Sections \ref{A-reprF} and \ref{S-CharIdentities} we provide the reader with an introduction 
to the algebra of quasicharacters and to the theory of characteristic identities.   In Section \ref{sec:SpinChainsProjectors} we
identify the irreducible quasicharacters with traces of representations of group elements (the $N$-fold Cartesian product $\times^N \SU(2)$), over totally coupled angular momentum states, labelled by binary coupling trees $T$ with $N$ leaves and associated intermediate labels. The angular momentum states themselves are realized via symmetrized tensor products of the fundamental spin-$\textstyle{\frac 12}$ representation. Explicit evaluations 
for certain low degree cases for $N=2$ and $N=3\,,4$ via projection techniques (Subsection \ref{subsec:ExplicitNeq2n3characters}) show that the quasicharacters are simply polynomials in 
$2 \times 2$ matrix traces of various group element strings. In Subsection  
\ref{subsec:ExplicitNeq2products}, certain pointwise products, of 
spin chain realizations for some generic $N=2$ irreducible quasicharacters (arbitrary spin $j$ coupled to spin-$\textstyle{\frac 12}$), multiplied by basic spin-$\textstyle{\frac 12}$ quasicharacters,
are themselves expanded in the quasicharacter basis, thus recovering certain structure constants
of the multiplicative ring. 

In Section \ref{Hamilton}\,, we briefly review the application of quasicharacter theory to Hamiltonian lattice quantum gauge theory, the central focus of our previous work which has motivated the present investigation.

To make the present paper more self-contained, in Appendix \ref{sec:IrrGchandRecoupling} we complement the constructive methods approach with a brief review of the abstract formalism of coupling and recoupling for quasicharacters (as developed in our previous work for both $\SU(2)$ 
\cite{Fuchs2018costratificationMR} and  compact $G$ \cite{jarvis2021quasicharactersMR}). We provide a summary of 
important properties of the quasicharacter calculus, such as the behaviour under change of coupling tree, and the role of Racah coefficients as structure constants under pointwise multiplication (Appendix \ref{subsec:Properties}).
In particular, the results of Subsection \ref{subsec:ExplicitNeq2products} for $N=2$ products can be viewed as a \emph{de novo} explicit evaluation of certain Racah $9j$ symbols. For completeness, the $9j$ result is given an independent proof (Appendix \ref{subsec:ExplicitNeq2Products9j})\,, where the general result, 
that the Racah coefficients which are multiplicative structure constants for arbitrary $N$\,, are 
expressible as products of $9j$ symbols, is also stated \cite{Fuchs2018costratificationMR,jarvis2021quasicharactersMR}\,. An index of notation used in the paper is given in table \ref{tab:NotationIndex}.

In the concluding Section \ref{sec:Concl}, we draw together some
further aspects of the concrete calculations of Section \ref{sec:SpinChainsProjectors}, 
with the results reviewed in the appendix, in relation to the combinatorial nature of the quasicharacter ring itself, and also formal role of the recoupling calculus in organizing its structure.



\section{The algebra of representative functions.}
\label{A-reprF}


For the convenience of the reader, let us recall some basics (we refer to \cite{Naimark} or \cite{Goodman} for more details\,).  
First, we will consider the general case of a compact Lie group. Next, we will limit our attention to the special case $G = \SU(2)$. 
We will assume that all representations under consideration are  continuous and unitary.

Let $G$ be a compact Lie group and let $\mf R(G)$ be the commutative algebra of representative functions of $G$.
Let $\widehat G$ be the set of isomorphism classes of finite-dimensional irreps of $G$. Given a finite-dimensional unitary representation $(H, \pi)$ of $G$, let $C(G)_\pi \subset \mf R(G)$ denote the subspace of representative functions {of $\pi$}, that is, the subspace spanned by all matrix 
coefficients $\langle \zeta , \pi(\cdot) v \rangle $ with $v \in H$ and $\zeta \in H^\ast$ of $\pi$. Moreover, 
let $\chibig_\pi \in C(G)_\pi$ be the character of $\pi$, defined by $\chibig_\pi(a) := \Tr\big(\pi(a)\big)$. The same notation will be used for the Lie group $G^N$.

The elements of $\widehat G$ will be labelled by the corresponding highest weight $\lambda$ relative to some chosen Cartan subalgebra and some chosen dominant Weyl chamber. Assume that for every $\lambda \in \widehat G$ a concrete unitary irrep $(H_\lambda,\pi_\lambda)$ of highest weight $\lambda$ in the Hilbert space $H_\lambda$ has been chosen. Given $\ul\lambda = (\lambda^1,\dots,\lambda^N) \in \widehat G^N$, we define a representation $(H_{\ul\lambda},\pi_{\ul\lambda})$ of $G^N$ by 
\beq
\label{G-irrepsGN}
H_{\ul\lambda} = \bigotimes_{i = 1}^N H_{\lambda^i}
\,,\quad 
\pi_{\ul\lambda}(\ul a) = \bigotimes_{i=1}^N \pi_{\lambda^i}(a_i)\,,
\eeq
where $\ul a = (a_1 , \dots , a_N)$. This representation is irreducible and we have 
$$
C(G^N)_{\pi_{\ul\lambda}} \cong \bigotimes_{i = 1}^N C(G)_{\pi_{\lambda^i}}
\,,
$$ 
isometrically with respect to the $L^2$-norms. Using this, together with the Peter-Weyl theorem for $G$, we obtain that $\bigoplus_{\ul\lambda \in \widehat G^N} C(G^N)_{\pi_{\ul\lambda}}$ is dense in $L^2 (G^N, {\rm d}^N a)$. Since $\bigoplus_{\ul\lambda \in \widehat G^N} C(G^N)_{\pi_{\ul\lambda}} \subset \bigoplus_{\pi \in \widehat{G^N}} C(G^N)_\pi$, this implies

\ble\label{L-ProdReps}

Every irreducible representation of $G^N$ is equivalent to a product representation $(H_{\ul\lambda},\pi_{\ul\lambda})$ with $\ul\lambda \in \widehat G^N$. If $(H_{\ul\lambda},\pi_{\ul\lambda})$ and $(H_{\ul\lambda'},\pi_{\ul\lambda'})$ are isomorphic, then $\ul\lambda = \ul\lambda'$.
\qed

\ele

Given $\ul\lambda \in \widehat G^N$, let $\pi^d_{\ul\lambda}$ denote the representation of $G$ on $H_{\ul\lambda}$ defined by  
\beq\label{pi-d}
\pi^d_{\ul\lambda}(a) := \pi_{\ul\lambda} (a, \ldots, a)
\,.
\eeq
This representation will be referred to as the diagonal representation induced by $\pi_{\ul\lambda}$. It is reducible and has the isotypical decomposition
$$
H_{\ul\lambda}
 =
\bigoplus_{\lambda \in \widehat G} H_{\ul\lambda,\lambda}
$$
into uniquely determined subspaces $H_{\ul\lambda,\lambda}$. Recall that these subspaces may be obtained as the images of the orthogonal projectors 
\beq
\label{Proj-irrep}
{\mathbb P}_\lambda
 := 
\dim (H_\lambda) \int_G \ol{\chibig_{\pi_\lambda}(a)} \, \pi_{\ul\lambda}(a) \, \mr d a 
\eeq
on $H_{\ul\lambda}$. These projectors commute with one another and with $\pi^d_{\ul\lambda}$. If an isotypical subspace $H_{\ul\lambda,\lambda}$ is reducible, we can further decompose it in a non-unique way into irreducible subspaces of isomorphism type $\lambda$. Let $m_{\ul\lambda}(\lambda)$ denote the number of these irreducible subspaces (the multiplicity of $\pi_\lambda$ in $\pi^d_{\ul\lambda}$) and let $\widehat G_{\ul\lambda}$ denote the subset of $\widehat G$ consisting of the highest weights $\lambda$ such that $m_{\ul\lambda}(\lambda) > 0$. In this way, we obtain a unitary $G$-representation isomorphism 
\beq\label{G-D-vp}
\vp_{\ul\lambda}
 : 
H_{\ul\lambda}
 ~\to~
\bigoplus_{\lambda \in \widehat G_{\ul\lambda}} 
 \,
\bigoplus_{k=1}^{m_{\ul\lambda}(\lambda)} 
H_\lambda
\,.
\eeq
Let 
\beq
\label{Proj-Inj}
\pr^{\ul\lambda}_{\lambda,k}
 : 
\bigoplus_{\lambda \in \widehat G_{\ul\lambda}} 
 \,
\bigoplus_{l=1}^{m_{\ul\lambda}(\lambda)} H_\lambda
 \to 
H_\lambda
 \,,\qquad
\mr i^{\ul\lambda}_{\lambda,k}
 : 
H_\lambda
 \to 
\bigoplus_{\lambda \in \widehat G_{\ul\lambda}} 
 \,
\bigoplus_{l=1}^{m_{\ul\lambda}(\lambda)} H_\lambda \, ,
\eeq
denote the natural projections and injections of the direct sum. For every $\lambda \in \widehat G_{\ul\lambda}$ and every $k,l = 1, \dots , m_{\ul\lambda}(\lambda)$, define a $G$-representation endomorphism $A^{\ul\lambda,\lambda}_{k,l}$ of $\pi^d_{\ul\lambda}$ by
\beq
\label{A-T}
A^{\ul\lambda,\lambda}_{k,l}
 := 
\frac{1}{\sqrt{\dim (H_\lambda)}}
 ~
\vp^{-1}_{\ul\lambda}
 \circ 
\mr i^{\ul\lambda}_{\lambda,k} \circ \pr^{\ul\lambda}_{\lambda,l}
 \circ 
\vp_{\ul\lambda}
\eeq
 \beq
\label{ReprF-Hom}
\BF{\ul\lambda}{\lambda}{k,l}(\ul a)
 := 
\sqrt{\dim(H_{\ul\lambda})}
\,
\Tr\left(\pi_{\ul\lambda}(\ul a) A^{\ul\lambda,\lambda}_{k,l}\right)
\,.
\eeq

\bsz\label{S-ReprF}

The family of functions 
$$
\left\{
\BF{\ul\lambda}{\lambda}{k,l}
~:~
\ul\lambda \in \widehat G^N
,~
\lambda \in \widehat G_{\ul\lambda}
\,,~
k,l = 1 , \dots , m_{\ul\lambda}(\lambda)
\right\}
$$
constitutes an orthonormal basis in  $L^2(G^N)^G$.
\qed
\esz
For the proof we refer to \cite{Fuchs2018costratificationMR}.  
{By this proposition, the $G$-invariant representative functions $\BF{\ul\lambda}{\lambda}{k,l}$, which in the sequel will be referred to as quasicharacters, span the subalgebra
$$
\mc R(G) : = \mf R(G^N)^G
$$
of $G$-invariant elements of $ \mf R(G^N)$.
}

{Next, let us turn to the discussion of the multiplicative structure  of  $\mc R(G)$. As the 
$\BF{\ul\lambda}{\lambda}{k,l}$ form a basis, it is enough to find the multiplication law for these functions.}
For that purpose, we assume that a unitary $G$-representation isomorphism $\vp_{\ul\lambda}$ as given by \eqref{G-D-vp} has been chosen for every $\ul\lambda \in \widehat G^N$ and every $N$. Denote
$d_{\lambda} := \dim H_\lambda$ and $d_{\ul \lambda} := \dim H_{\ul \lambda}$. 
Then, using \eqref{ReprF-Hom}, we can write 
 \al{\nonumber
\BF{\ul\lambda_1}{\lambda_1}{k_1,l_1}(\ul a)&
 \,
\BF{\ul\lambda_2}{\lambda_2}{k_2,l_2}(\ul a)
\\ \label{G-MF-1}
 & =
\sqrt{d_{\ul\lambda_1} d_{\ul\lambda_2}}
 \,
\Tr
 \left(
 \left(
A^{\ul\lambda_1,\lambda_1}_{k_1,l_1} \otimes A^{\ul\lambda_2,\lambda_2}_{k_2,l_2}
 \right)
 \circ
\Big(\pi_{\ul\lambda_1}(\ul a) \otimes \pi_{\ul\lambda_2}(\ul a)\Big)
 \right)
\,.
 }
If we wish to decompose this product in terms of the $\BF{\ul\lambda}{\lambda}{k,l}$, we have to decompose 
the $G^N$-representation $\pi_{\ul\lambda_1} \otimes \pi_{\ul\lambda_2}$ into $G^N$-irreps $\ul\lambda$ and then relate the latter
to the basis functions using $\vp_{\ul\lambda}$. This procedure is explained in detail in 
\cite{Fuchs2018costratificationMR} with the result stated there as  Proposition 3.10.
%

If we introduce composite indices $I, J\,, \cdots$ representing the label sets $\ul \lambda, \lambda, k, l$,
the product
rule for the quasicharacter ring reads schematically as follows:
\begin{align}
\label{eq:QuasiMultRule}
\chibig_{I_1} \chibig_{I_2} = \left.\sum\right._{I_3} C^{I_3}_{I_1I_2} \chibig_{I_3}\, , 
\end{align}
with the structure constants $C^{I_3}_{I_1I_2}$ depending on two auxiliary unitary $G$-representation isomorphisms constructed by using the chosen isomorphism $\vp_{\ul\lambda}$ and are given, up to normalization, by group theoretical Racah recoupling coefficients for $G$. 
Below, we will see that for $G = \SU(2)$ the coefficients boil down to recoupling coefficients of angular momentum theory.

Thus, let us turn to the special case $G = \SU(2)$. Here, the highest weights of irreps $\widehat \SU(2)$ are in one-to-one correspondence with  spins $j = 0 , \frac 1 2 , 1 , \frac 3 2 , \dots$. We will use the common notation $D_j$ for $\pi_j$. Thus, $(H_j,D_j)$ is the standard $\SU(2)$-irrep of spin $j$, with dimension $d_j = 2j+1 (\equiv {[}j{]})$ spanned by the orthonormal ladder basis $\{\ket{j,m} : m = -j , -j+1 , \dots , j\}$ which is unique up to a phase. Accordingly, every  sequence of highest weights corresponds to a sequence $\ul j = 
j_1,j_2, \cdots, j_N$ of spins. We write \smash{$D_{\ul j} \equiv \pi_{\ul j}$} for the corresponding irrep of $\times^N \SU(2) =\SU(2)^N$ and $D^d_{\ul j} \equiv \pi^d_{\ul j}$ for the induced diagonal representation of $\SU(2)$. To fix the $G$-representation isomorphisms
\beq\label{G-vp1j}
\vp_{\ul j}
 : 
H_{\ul j}
 \to
\bigoplus_j \bigoplus_{i=1}^{m_{\ul j}(j)} H_j
\,,
\eeq
we choose the following reduction scheme for tensor products of $N$ irreps of $\SU(2)$. Given nonnegative half integers $s_1$, $s_2$, denote
$$
\langle s_1 , s_2 \rangle
 := 
\{|s_1-s_2| , |s_1-s_2|+1 , |s_1-s_2|+2 , \dots , s_1+s_2\}
$$
and recall that the representation space $H_{s_1} \otimes H_{s_2}$ decomposes into unique irreducible subspaces $(H_{s_1} \otimes H_{s_2})_s$ of spin $s \in \langle s_1 , s_2 \rangle$. We start by decomposing $H_{j_1}\otimes H_{j_2}$ into the unique irreducible subspaces $(H_{j_1}\otimes H_{j_2})_{l_2}$ with $l_2 \in \langle j_1 , j_2 \rangle$. Then, we decompose the invariant subspaces 
$$
(H_{j_1}\otimes H_{j_2})_{l_2} \otimes H_{j_3}
 \subset 
H_{j_1}\otimes H_{j_2}\otimes H_{j_3}
$$
into unique irreducible subspaces 
$$
((H_{j_1}\otimes H_{j_2})_{l_2} \otimes H_{j_3})_{l_3}
 \,,\quad
l_3 \in \langle l_2 , j_3 \rangle
\,.
$$
Iterating this, we end up with a decomposition of $H_{\ul j}$ into unique irreducible subspaces 
\beq\label{G-D-Hjl}
H_{\ul j , \ul l}
 :=
(
 \cdots
((H_{j_1}\otimes H_{j_2})_{l_2} \otimes H_{j_3})_{l_3} 
 \cdots \otimes 
H_{j_N})_{l_N}
\,,
\eeq
where $\ul l = (l_1 , \dots , l_N)$ is a sequence of nonnegative half integers satisfying $l_1 = j_1$ and $l_i \in \langle l_{i-1} , j_i\rangle$ for $i = 2 , 3 , \dots , N$. 
Note that for each $l_i$, the sequence $l_i, j_{i+ 1}, \cdots, j_N$ labels
a subspace $H_{l_i}\otimes H_{j_{i+1}}\otimes \cdots\otimes H_{j_N}$ isomorphic to an irreducible representation of member $N- i + 1$ of a descending chain of subgroups
\beq
\label{eq:SubgroupChaindDef}
\times^N \SU(2) \supset \times^{N-1}\SU(2)\supset\cdots \times^2 \SU(2)\supset \SU(2)^d\,.
\eeq
Let us denote the totality of such sequences by $R_{\ul j}$. Moreover, denote
\beq
\label{eq:RjDef}
R_{\ul j} (j) = \{\ul l \in R_{\ul j} : l_N = j\}
\,.
\eeq
Then, $m_{\ul j}(j) = |R_{\ul j} (j)|$\,,
and the isotypical component $H_{\ul j,j}$ of $H_{\ul j}$ is given by the direct sum of the subspaces $H_{\ul j , \ul l}$ with $\ul l \in R_{\ul j}(j)$.

To define the isomorphism $\vp_{\ul j}$, we choose\footnote{Any other choice would yield the same basis vectors but multiplied by a phase which depends on $\ul l$ only.} ladder bases in the irreducible subspaces $H_{\ul j , \ul l}$. Denote their elements by $\ket{\ul j , \ul l , m}$, where $m = - j , -j+1 , \dots , j$. Then,
\beq
\label{eq:uljullmCoupled}
\{\ket{\ul j , \ul l , m} : \ul l \in R_{\ul j} (j), m = - j , -j+1 , \dots , j\}
\eeq
is an orthonormal basis in $H_{\ul j}$, 
and we can use the sequences $\ul l \in R_{\ul j}(j)$ to label the copies of $H_j$ in the direct sum decomposition of the target space of $\vp_{\ul j}$. Thus,  the natural injections \eqref{Proj-Inj} related to this decomposition read $\mr i^{\ul j}_{j,\ul l}$, the basis functions read $\BF{\ul j}{j}{\ul l,\ul l'}$ and the endomorphisms appearing in their definition read $A^{\ul j,j}_{\ul l,\ul l'}$. We define $\vp_{\ul j}$ by 
$$
\vp_{\ul j}(\ket{\ul j,\ul l,m}) := \mr i^{\ul j}_{j,\ul l}(\ket{j,m})
\,,
$$
where $\ket{j,m}$ denotes the elements of the orthonormal ladder basis in $H_j$. Under this choice, one obtains
\beq\label{Form-A}
A^{\ul j,j}_{\ul l,\ul l'}
 =
\frac{1}{\sqrt{d_j}} \, \sum_{m=-j}^j \ket{\ul j,\ul l,m} \bra{\ul j,\ul l',m}
 \,,\qquad
\ul l,\ul l' \in R(\ul j,j) \, ,
\eeq
and
\beq\label{Form-chi}
\BF{\ul j}{j}{\ul l,\ul l'}(\ul a)
 =
\sqrt{\frac{d_{\ul j}}{d_j}} \, \sum_{m=-j}^j 
\bra{\ul j,\ul l',m} D^{\ul j}(\ul a) \ket{\ul j,\ul l,m}
 \,,\qquad
\ul l,\ul l' \in R(\ul j,j)
\,.
\eeq
Using the matrix elements $D^{j_i}_{m_i,m_i'}$, $i=1,\dots,N$, together with the Clebsch-Gordan coefficients  
$C^{s_1,s_2,s}_{m_1,m_2,m}$, we obtain the following. 

\bsz\label{S-chi-D}

We have
$$
\BF{\ul j}{j}{\ul l,\ul l'}(\ul a)
 =
\sqrt{\frac{d_{\ul j}}{d_j}}
\sum_{m=-j}^j \sum_{\ul m} \sum_{\ul m'}
 \,
C(\ul j,\ul l,\ul m) \, C(\ul j,\ul l',\ul m')
 \,
D^{j_1}_{m_1',m_1}(a_1) \cdots D^{j_N}_{m_N',m_N}(a_N)
\,,
$$
where $\sum_{\ul m}$ means the sum over all sequences $\ul m = (m_1 , \dots , m_N)$ such that 
$$
m_i = - j^i , \dots , j^i \text{ for } i = 1 , \dots , N
 \,,\qquad
m_1 + \cdots + m_N = m
\,,
$$
and where
$$
C(\ul j,\ul l,\ul m)
 =
C^{j^1,j^2,l^2}_{m_1,m_2,m_1+m_2} 
C^{l^2,j^3,l^3}_{m_1+m_2,m_3,m_1+m_2+m_3}
 \cdots
C^{l^{N-1},j^N,l^N}_{m_1+\cdots+m_{N-1},m_N,m}
\,.
$$
\qed
\esz

For the proof we refer to \cite{Fuchs2018costratificationMR}.
Next, one can derive the following multiplication law for the representative functions. 

\bsz\label{Gen-MultLaw}

For $G = \SU(2)$, the multiplication law for the basis functions $\BF{\ul j}{j}{\ul l,\ul l'}$ reads as follows:
 \al{\nonumber
\BF{\ul j_1}{j_1}{\ul l_1,\ul l_1'} \cdot \BF{\ul j_2}{j_2}{\ul l_2,\ul l_2'} 
 = &
\sqrt{\frac{d_{\ul{j_1}}d_{\ul{{j_2}}}}{d_{j_1}d_{j_2}}}
 ~
\sum_{\ul j \in \langle \ul j_1,\ul j_2 \rangle}
	~
\sum_{\ul l,\ul l'\in R_{\ul j}( j)}
 ~
\sqrt{\frac{d_j}{d_{\ul j}}}
\\ \nonumber
& \hspace{1cm}
U_{\ul j_1,\ul j_2}(\ul j,\ul l;\ul l_1,\ul l_2)
 \,
U_{\ul j_1,\ul j_2}(\ul j,\ul l';\ul l_1',\ul l_2')
 \,
\BF{\ul j}{j}{\ul l,\ul l'}
\,,
 }
where 
\beq\label{G-U-SU2}
U_{\ul j_1,\ul j_2}(\ul j,\ul l;\ul l_1,\ul l_2)
 =
\braket{\ul j_1,\ul j_2;\ul j,\ul l,m}{\ul j_1,\ul j_2;\ul l_1,\ul l_2;j,m}
\eeq
for every $\ul j \in \langle \ul j_1,\ul j_2 \rangle$, $j \in \langle j_1,j_2 \rangle$ and $\ul l \in R(\ul j, j)$, and for any admissible $m$.
\qed
\esz
For the proof see \cite{Fuchs2018costratificationMR}. It can be seen from the proof that the recoupling coefficients $U$ are (up to some phase choices) uniquely determined by the above choice of the isomorphism $\vp_{\ul\lambda}$. Up to normalization, they are given by what is known as $3(2N-1)j$ symbols.
Finally, it turns out that the recoupling coefficients can be expressed in terms of the recoupling coefficients for $N=2$. 

\ble\label{L-symbols}

The recoupling coefficients $U_{\ul j_1,\ul j_2}(\ul j,\ul l;\ul l_1,\ul l_2)$ are given by
$$
U_{\ul j_1,\ul j_2}(\ul j,\ul l;\ul l_1,\ul l_2)
 =
\prod_{i=2}^N
\begin{pmatrix}
l_1^{i-1} & l_2^{i-1} & l^{i-1}
\\
j_1^i & j_2^i & j^i
\\
l_1^i & l_2^i & l^i
\end{pmatrix}
,
$$
where $l_1^1 := j_1^1$, $l_2^1 := j_2^1$ and $l^1 := j^1$. Here,  
$$
\begin{pmatrix}
j_1 & j_2 & j_3
\\
j_4 & j_5 & j_6
\\
j_7 & j_8 & j_9
\end{pmatrix}
 =
\sqrt{d_{j_3}d_{j_6}d_{j_7}d_{j_8}}
\begin{Bmatrix}
j_1 & j_2 & j_3
\\
j_4 & j_5 & j_6
\\
j_7 & j_8 & j_9
\end{Bmatrix}, 
$$
with the bracket on the right hand side denoting the $9$j Racah symbols.
\ele

To summarize, the above results relate the structure of the algebra $\mathcal R(\SU(2))$ to the combinatorics of recoupling theory of angular momentum as provided in \cite{biedenharn1981angular,biedenharn1981racah,louck2008unitary,Yutsis,wormer2006angular}.
Moreover, as  there exist efficient calculators for $9j$ symbols, provided e.g. by the Python library \texttt{SymPy} \cite{meurer2017sympy} or online by Anthony Stone's Wigner coefficient calculator\footnote{See http://www-stone.ch.cam.ac.uk/cgi-bin/wigner.cgi}, Lemma \rref{L-symbols} provides an explicit knowledge of the multiplication law in the commutative algebra 
$\mc R(\SU(2))$ .


\section{Characteristic identities. The projection operator method.}
\label{S-CharIdentities}


Recall that the analysis of {the algebra $\mc R(\SU(2))$ } in terms of generalized Clebsch-Gordan coefficients provided in the previous section relies on the sequence of irreducible modules displayed in (\ref{G-D-Hjl}). Clearly, associated with that decomposition of 
$H_{\ul j}$, there is an ordered composition of projection operators from pairwise tensor products of irreducible spin representations on to specific irreducible constituents.  This observation provides the algorithm that we are going to propose as an alternative 
calculational method for dealing with {the algebra $\mc R(\SU(2))$}. It will be based
on the general theory of characteristic identities for Lie algebras (see \cite{green1971characteristic,bracken1971vector}, \cite{OBrien1977}, \cite{gould1978tensor}, \cite{jarvis1979casimir},  \cite{gould1985characteristic,isaac2015characteristic}\,). 

For the convenience of the reader, let us give a short introduction to the subject. Our presentation is along the lines of 
\cite{OBrien1977} and \cite{gould1978tensor} (see also \cite{kostant1975tensor}). We also refer to \cite{humphreys1980introduction} for the basics on Lie algebras.

In the standard notation of \cite{humphreys1980introduction}, let $L$ be a semi-simple Lie algebra (over some field $\F$), let $H$ be a fixed Cartan subalgebra with dual $H^\ast$. Let $\Phi$ be the set of roots relative to $H$ and let $ \Phi^+$ be the set of positive roots. Let us 
choose a basis $\{x_1, \ldots,x_l \} $ of $L$ and let $\{x^1, \ldots,x^l \} $ be the dual basis with respect to the Killing form of $L$. 
In particular, we can choose a basis consisting of root space elements, together with a basis $\{h_1, \ldots, h_m \}$ of $H$.
Moreover, let us denote $\delta = \tfrac{1}{2} \sum_{\alpha \in \Phi^+} \alpha$, let $(\cdot, \cdot)$ be the inner product on $H^\ast$ induced by the symmetric Killing form and $\Lambda$ be the labelling operator, that is, the operator which by acting on an irrep of $L$ with highest weight $\mu$ coincides with $\mu$. For a chosen basis $\{h_1, \ldots, h_m \}$ of $H$ it is represented by a vector operator with components $\Lambda(h_i)$ taking constant values $\mu(h_i)$.

Now, let $\pi$ be a finite-dimensional representation of $L$ and let us consider a matrix operator $A$ with entries
\beq
\label{MatrOp-A}
A^i{}_j = -\frac{1}{2} \sum_{r=1}^l \left( \pi(x_r)^i{}_j x^r + \pi(x^r)^i{}_j x_r \right)\, ,
\eeq
together with its adjoint $\overline{A}^i{}j = - A_j{}^i $. Let $T$ be a contragredient tensor operator relative to $\pi$. Then 
one can show the following identities:
\beq
\label{Id-A-C-L}
A T - T \overline{A} =  (2A -\pi(c_L) ) T = [c_L, T] \, ,
\eeq
where $c_L$ is the universal Casimir element of $L$. 

Let $(V_\lambda, \pi_\lambda)$ be a finite-dimensional  irreducible $L$-module with highest weight $\lambda$, and for this section let $\{\lambda_i\}_{i=1}^k$  
$= \{\lambda_1, \ldots, \lambda_k\}$ be the list of the distinct weights in $V(\lambda)$. Let $A^\lambda$ be the corresponding matrix operator 
given by \eqref{MatrOp-A} with $\pi$ replaced by $\pi_\lambda$. Then, 
any irreducible  contragredient tensor operator $T$ with highest weight $\lambda$ may be decomposed into shift tensors $T_{[i]}$, $i=1,\cdots, k$\,,
which decrease the eigenvalue of the labelling operator $\Lambda$ on an irrep by the weight $\lambda_i$, 
$$
\Lambda(h_j) T_{[i]} = T_{[i]} (\Lambda - \lambda_i) (h_j) \, .
$$
In terms of the shift operators, the identities \eqref{Id-A-C-L} take the following form:
\beq
\label{Id-A-C-lambda}
( A^\lambda - \Lambda_i) T_{[i]} = 0 \, , \quad T_{[i]} (\overline{A}^\lambda - \overline{\Lambda}_i )= 0\, , 
\eeq
where $\Lambda_i$ and $\overline{\Lambda}_i$ are given by 
\begin{align}
\label{PolId-A}
\Lambda_i  &= \tfrac{1}{2} (\lambda, \lambda + 2 \delta)  - \tfrac{1}{2} (\lambda_i, 2(\Lambda + \delta) + \lambda_i) \, ,
\\
\label{PolId-ovA}
\overline{\Lambda}_i & = \tfrac{1}{2} (\lambda, \lambda + 2 \delta) + \tfrac{1}{2} (\lambda_i, 2(\Lambda + \delta) - \lambda_i) \, .
\end{align}
Moreover, we have the following.
\bsz
The matrix operators $A^\lambda$ and $\overline A^\lambda$ satisfy the 
following characteristic identities: 
\beq
\label{CharIdent-A-ovA}
\prod_{i = 1}^k \left( A^\lambda -\Lambda_i \right) = 0 \, , \quad 
\prod_{i = 1}^k \left( \overline{A}^\lambda - \overline{\Lambda}_i \right) = 0 \, . 
\eeq
\esz

\begin{rem}
\label{A-lambda-mu}
If $V(\mu)$ is an irreducible module, then $A^\lambda$ acting on $V(\mu)$ yields
\beq
\label{Rel-A-CL}
A^{\lambda, \mu} =  \tfrac{1}{2} \left(\pi_\lambda \otimes \pi_\mu (c_L) - \pi_\lambda( c_L) \otimes 1 - 1 \otimes  \pi_\mu( c_L)\right)  \, .
\eeq
Thus, for every highest weight module $V_\mu$, $A^\lambda$ yields an operator $A^{\lambda, \mu}$ acting on the tensor product $V(\lambda) \otimes V(\mu)$. If we decompose this product into irreps,
\beq
\label{Clebsch-Gordan-Dec}
V(\lambda) \otimes V(\mu) = \bigoplus_{i = 1}^k m_{\lambda \mu}^{\mu + \lambda_i} \, V(\mu + \lambda_i ) \, ,
\eeq
then $A^{\lambda, \mu}$ yields the constant value
\beq
\label{Eigenvalue-lambda-mu}
\alpha_i = \tfrac{1}{2} (\lambda, \lambda + 2 \delta)  - \tfrac{1}{2} (\lambda_i, 2(\mu + \delta) + \lambda_i) 
\eeq
on each subspace $V(\mu + \lambda_i ) $. This, together with the observation that a diagonal matrix $B$ with distinct eigenvalues 
$b_i$ satisfies $\Pi_{i = 1}^k (B - b_i) = 0$, essentially yields the proof of the characteristic identities \eqref{CharIdent-A-ovA}.
\qeb
\end{rem}
By extending the relation $A^\lambda T_{[i]} = \Lambda_i T_{[i]}$ in \eqref{Id-A-C-lambda} to any polynomial $p(A^\lambda)$ and choosing, in particular, $p(A^\lambda) = \Pi_{l \neq j } (A^\lambda - \Lambda_l)$, we obtain
\beq
\label{Id-A-C-lambda-p}
 \Pi_{l \neq j } (A^\lambda - \Lambda_l) T_{[i]} = \Pi_{l \neq j } (\Lambda_i - \Lambda_l) T_{[i]} \, .
\eeq
Thus, we have 
\beq
\label{Id-A-C-lambda-p-Proj}
 \PK_j^\lambda T_{[i]} = \delta_{ij} T_{[i]}  \, ,
\eeq
where $\PK_j^\lambda$ is a projection operator given by 
\beq
\label{ProjOp-charId}
\PK_j^\lambda = \prod_{l \neq j} \frac{A^\lambda- \Lambda_l}{\Lambda_j - \Lambda_l} \, .
\eeq
Correspondingly, for $\overline{A}^\lambda$, we obtain
\beq
\label{ProjOp-charId-ov} 
T_{[i]} \overline{ \PK}_j^\lambda = \delta_{ij} T_{[i]} \, , \quad 
\overline{\PK}_j^\lambda = \prod_{l \neq j} \frac{\overline{ A}^\lambda -  \overline{\Lambda}_l}{\overline{\Lambda}_j - \overline{\Lambda}_l} \, .
\eeq
If we decompose the tensor operator $T$ into its shift components, $T = \sum_{i = 1}^k T_{[i]} $, then we clearly have
\beq
\label{Dec-ShifOp}
T_{[i]} = \PK_i^\lambda T \, , \quad T_{[i]} = T \overline{\PK}_i^\lambda \, .
\eeq
Now, equations \eqref{CharIdent-A-ovA}, \eqref{ProjOp-charId} -- \eqref{Dec-ShifOp} immediately imply the following. 
\bsz
\label{S-PropProjOp}
The projection operators $\PK_j^\lambda$ and $\overline{\PK}_j^\lambda$ fulfil
\begin{align}
\label{PropProjOp-1}
A^\lambda \PK_j^\lambda  = \Lambda_j \PK_j^\lambda \, , \quad \overline{A}^\lambda \overline{\PK}_j^\lambda = \overline{\Lambda}_j \overline {\PK}_j \, ,\\
\label{PropProjOp-2}
\PK_i^\lambda \PK_j^\lambda  = \delta_{ij} \PK_j^\lambda \, , \quad 
\overline{\PK}_i^\lambda \overline{\PK}_j^\lambda  = \delta_{ij}  \overline{\PK}_j^\lambda \, , \\
\label{PropProjOp-3}
\sum_{i = 1}^k \PK_i^\lambda  = \II \, , \quad \sum_{i = 1}^k \overline{\PK}_i^\lambda = \II \, .
\end{align}
\esz
Note that \eqref{PropProjOp-1}  yields the spectral decompositions of $A^\lambda$ and $\overline{A}^\lambda $. That is, given the Clebsch-Gordan decomposition \eqref{Clebsch-Gordan-Dec}, the operator $\PK_j^\lambda$ projects onto the irreducible 
component $ V(\mu + \lambda_j)$ and, thus,  the action of $A^\lambda \PK_j^\lambda$ onto an element of 
$V(\lambda) \otimes V(\mu)$ yields a vector of $V(\mu + \lambda_j)$ multiplied by the eigenvalue $\alpha_j$ given by 
\eqref{Eigenvalue-lambda-mu}. In the sequel, if an irreducible module $V_\mu$ is chosen, then in accordance with the 
notation $A^{\lambda, \mu}$, the corresponding projectors will be denoted by
\beq
\label{Proj-lambda-mu}
\PK_{\lambda, \mu}^{\mu + \lambda_j} \, .
\eeq

\begin{rem}
\label{B-Reduction}
In general, for a given irrep $V(\mu)$, not all of the representations $V(\mu + \lambda_i)$ occur in the 
Clebsch-Gordan decomposition \eqref{Clebsch-Gordan-Dec}. This fact leads to a reduction of the characteristic identities 
\eqref{CharIdent-A-ovA}. As was shown in \cite{gould1978tensor}, the product over all $i$ in \eqref{CharIdent-A-ovA} gets reduced to the 
following subset:
$$
\overline{I} (\lambda, \mu) 
= \left\{ i : (\mu + \delta - \lambda_i, \alpha ) \neq 0  \, \, \text{for all } \, \, \alpha \in \Phi^+ \right\} \, .
$$
\qeb
\end{rem}

Turning to the formalism for the case $G=\SU(2)$, we denote the matrix operators $A^{\lambda, \mu}$ by $\bJ^{j,k}$ and the corresponding projectors by ${\mathbb P}_{j,k}^{k  + \lambda_j}$. For convenience, from now on, we will use the Dynkin labelling, that is, we will write $j = 2 \tfrac 12 j$ for spin $\tfrac 12 j$. In this notation,  the Casimir invariant  reads $\textstyle{\frac 14}j(j+2)$ ) (c.f. equation (\ref{eq:Casimir}))\,.

From \eqref{Rel-A-CL} we have 
\beq
\label{Rel-A-CL-1}
\bJ^{j,k} =  \tfrac{1}{2} \left(\pi_j \otimes \pi_k (c_L) - \pi_j( c_L) \otimes 1 - 1 \otimes  \pi_k( c_L)\right)  \, .
\eeq
Note that the angular momentum generators for arbitrary spin are 
$\bfJ^j = \pi_j(\tfrac 12 \boldsig)$, so that 
\beq
\label{Rel-A-CL-2}
\bJ^{j,k}=({\bfJ^j}\otimes {\sf Id})\cdot ({\sf Id}\otimes {\bfJ^k})
\equiv {\bfJ^j} \cdot {\bfJ^k} \, , 
\eeq
with eigenvalues
\beq
\label{eq:MathbbJjkDef}
\alpha_{\ell} = \textstyle{\frac 12}\big(\textstyle{\frac 12}\ell(\textstyle{\frac 12}\ell+1)
- \textstyle{\frac 12}j(\textstyle{\frac 12}j+1)-\textstyle{\frac 12}k(\textstyle{\frac 12}k+1)\big)\,,\quad \mbox{for} \quad\ell \in \langle j,k\rangle\,.
\eeq
In the sequel, the following projectors will be useful.

\ble[Angular momentum projectors for $H_1 \otimes H_j$ and $H_2 \otimes H_j$]
\label{lem:SU2Projs}\mbox{}\\
\beq
\label{eq:stretchP}
{\mathbb P}^{j\!+\!1}_{(1,j)} := \,  \frac{  2{\mathbb J}^{(1,j)} \!+\! (\hj\!+\!1) \unit \otimes \unit^j}{j\!+\!1} \, , \quad 
{\mathbb P}^{j-1}_{(1,j)} := \frac{\hj\unit\otimes \unit^j -2{\mathbb J}^{(1,j)}}{j\!+\!1}\, .
\eeq

\begin{align}
\label{eq:Projectorsj2}
{\mathbb P}^{j\!+\!2}_{(2,j)} := &\,  \frac{(\bJ^{2,j})^2\!+\!(\hj\!+\!2)(\bJ^{2,j}) \!+\! (\hj\!+\!1)\unit^{2}\otimes \unit^j}{(\hj\!+\!1)(j\!+\!1)}\, , \nonumber \\
{\mathbb P}^{j}_{(2,j)} := &\,  \frac{-(\bJ^{2,j})^2-(\bJ^{2,j}) \!+\! \hj(\hj\!+\!1)
\unit^{2}\otimes \unit^j}{(\hj\!+\!1)\hj}\, , \nonumber \\
{\mathbb P}^{j-2}_{(2,j)} := &\,  \frac{(\bJ^{2,j})^2\!+\!(1-\hj)(\bJ^{2,j}) -\hj \unit^{2}\otimes \unit^j}{(j\!+\!1)\hj}\, .  
\end{align}
\bx
\ele

\bpf
We will prove the first formula, to check the remaining ones is left to the reader.
By \eqref{ProjOp-charId}, we have 
$$
\PK^{j+1}_{(1,j)} =  \frac{\bJ^{(1,j)} -\alpha_{j-1} \II}{\alpha_{j+1} - \alpha_{j-1}}\, .
$$
Next, by \eqref{Rel-A-CL-2}, we have the eigenvalues $\alpha_{j+1} = \tfrac{1}{4}j $ and
 $\alpha_{j-1} = -\tfrac{1}{4}j - \frac{1}{2}$. Inserting them into the above formula 
yields the assertion. 
\epf




\section{Spin chain techniques for $\SU(2)$ quasicharacters.}
\label{sec:SpinChainsProjectors}

In this section we are going to study the algebra of $\SU(2)$ quasicharacters in terms of the characteristic projectors introduced above: 
\ben
\item 
We will develop an algebraic calculus which allows to determine the quasicharacters in terms of trace invariants. 
\item 
We will apply our method to the explicit analysis of the multiplication law in terms of trace invariants. 
\een
Clearly, one cannot hope to produce general formulae, so in this paper we will concentrate on the cases $N = 2$ as well as examples from
$N =3, N=4$. Our presentation will be expository rather than formal, with the algorithmic steps being presented by way of examples. 

To start with, recall from \eqref{G-D-Hjl} that in terms of the chosen reduction scheme we have a decomposition of $H_{\ul j}$  into irreducible subspaces given by 
\beq
\label{G-D-Hjl-1}
H_{\ul j , \ul l}
 :=
(
 \cdots
((H_{j^1}\otimes H_{j^2})_{l^2} \otimes H_{j_3})_{l^3} 
 \cdots \otimes 
H_{j^N})_{l^N}
\, .
\eeq
There is an associated family of operators projecting onto these irreps given by 
\beq
\label{Dec-ProjOp}
{\mathbb P}^j_{\ul j, \ul l} = 
{\mathbb P}^j_{(l_{N-1} , j_N)}\circ({\mathbb P}^{l_{N-1}}_{l_{N-2}, j_{N-1}} \otimes \II ) 
\cdots ({\mathbb P}^{l_3}_{l_2, j_3} \otimes \II )  \circ 
({\mathbb P}^{l_2}_{j_1, j_2} \otimes \II ) \, ,
\eeq
with $l_N = j$. If we choose ladder bases $ \ket{\ul j,\ul l,m} $ in the irreducible subspaces $H_{\ul j , \ul l}$, then we obtain
\beq
\label{Proj-ladderB}
{\mathbb P}^j_{\ul j, \ul l} =  \sum_{m=-j}^j \ket{\ul j,\ul l,m}   \bra{\ul j,\ul l,m} \, .
\eeq
Thus, we can rewrite \eqref{Form-chi} as follows:
\begin{align*}
\BF{\ul j}{j}{\ul l,\ul l'}(\ul u)
& = \sum_{m=-j}^j \bra{\ul j,\ul l',m} D^{\ul j}(\ul u) \ket{\ul j,\ul l,m} \\
& = \sum_{m=-j}^j \bra{\ul j,\ul l',m}  \left( \left.\bigoplus\right._{j,\ti {\ul l}} {\mathbb P}^j_{\ul j, \ti {\ul l}} \right)  D^{\ul j}(\ul u) 
 \left( \left. \bigoplus \right._{j,{\ti {\ul l}}^\prime}  {\mathbb P}^j_{\ul j, {\ti {\ul l}}^\prime  } \right)  \ket{\ul j,\ul l,m} \\
& = \sum_{m=-j}^j \bra{\ul j,\ul l',m}  {\mathbb P}^j_{\ul j, \ul l'}  D^{\ul j}(\ul u) 
 {\mathbb P}^j_{\ul j, \ul l }  \ket{\ul j,\ul l,m}  \,.
\end{align*}
Here, $\ket{\ul j,\ul l,m} \in H_{\ul j , \ul l}$ and $  \ket{\ul j,\ul l^\prime,m} \in  H_{\ul j , \ul l^\prime}$, where 
$l_N = l^\prime_N = j$. Thus, $H_{\ul j , \ul l} \cong H_{\ul j , \ul l^\prime}$  and both spaces may be identified with 
the abstract irrep $H_j $ with ladder basis $\ket{ j m}$. Under this isomorphism the sum in the above formula may be  
interpreted as the trace over $H_j$ and we obtain
\beq
\label{Quasich-Proj-gen}
\BF{\ul j}{j}{\ul l,\ul l'}(\ul u) =  \mathrm {Tr} \left( {\mathbb P}^j_{\ul j, \ul l'}  D^{\ul j}(\ul u) 
 {\mathbb P}^j_{\ul j, \ul l } \right) \, .
\eeq

\begin{rem}[Notation]
\label{B-Notation}
\mbox{}\\
In the sequel, we use the following simplified notation. First, recall that we adopt Dynkin notation and from now 
we will write $\pi_j (\cdot)$ for an irrep with spin $\tfrac{1}{2} j$ and dimension $j +1$. Moreover, we will write
$$
\BF{\ul j}{j}{\ul l,\ul l'} = \chibig^j_{(\ul j), \ul l,\ul l'} \, .
$$
Group elements $\ul u$ , and generically local operators (on single members of tensor product spaces) are denoted $u$, $v$, $a$, $\II$, or $\sf x, \sf y, \ldots$;  nonlocal operators are capitalized in blackboard font, e.g. 
${\mathbb J}^{j,k}$, ${\mathbb P}^{j,1}$ or ${\mathbf J}^1 \cdot {\mathbf J}^2 $ where the separate spaces are indicated
explicitly.  Repeated tensor products are written as
$u^{\otimes_j}$, $\unit^{\otimes_j}\equiv \unit^j$ (grouplike), ${\sf x}^{\otimes_j}$ (primitive).
The forms $\left.\sum\right._{\sf x} \!\left.\sum\right._{\sf y} {\sf x} \cdots {\sf y}\cdots   {\sf y'}\cdots {\sf x'}$
will signify correlated sums over independent basis sets, 
inserted at various positions in tensor products. 
\qeb
\end{rem}
In this notation,  for the angular momentum cases  $N=2,3$,  we have 
\begin{align}\label{eq:Neq23quasidefs}
{\chibig}^l_{jk}(u,v) =&\, \Tr\big( {\mathbb P}^l_{j,k} \pi_j(u)\otimes \pi_k(v)\big)\,; \\
{\chibig}^j_{(j_1 j_2) j_3,k k'}(u,v,w) =&\, \Tr\big( {\mathbb P}^j_{(j_1j_2)j_3,k}  
\pi_{j_1}(u)\otimes \pi_{j_2}(v)\otimes \pi_{j_3}(v)  {\mathbb P}^j_{(j_1j_2)j_3,k'} \big)\, ,
\end{align}
with the understanding that in the lists $\ul l, \ul l^\prime$ the final $j$ is omitted. 

Now, having fixed the notation, let us explain the basic idea of our calculus. For that purpose, 
recall the following classical result of Cayley and Sylvester. Let $H_1 \cong \C^2$ be the basic $2$-dimensional representation space 
of $\SU(2)$. Then, every irreducible representation of $\SU(2)$ is isomorphic to $S^n(H_1)$ for some $n$, where $S^n(H_1)$ is 
the symmetric power of $H_1$. Thus, in order to realize the irreducible representation $\pi_j(u)$
we use the symmetrized tensor product of $\otimes^j \pi_1(u)\equiv \otimes^j u$\,, with generator
\[
\pi_j({\mathbf J}) =\textstyle{\frac 12}{\mathbb S}^j \mbox{\boldmath{$\sigma$}}^{\otimes_j}{\mathbb S}^j \, ,
\] 
where  $\mbox{\boldmath{$\sigma$}}^{\otimes_j}$ is the $(j-1)$-fold iterated coproduct
\[
\mbox{\boldmath{$\sigma$}}^{\otimes_j} :=
\mbox{\boldmath{$\sigma$}} \otimes \unit \otimes \cdots \otimes \unit +
\unit \otimes \mbox{\boldmath{$\sigma$}} \otimes \unit \otimes \cdots \otimes \unit
\cdots +
\unit \otimes \unit \otimes \cdots \otimes \mbox{\boldmath{$\sigma$}}
\]
and ${\mathbb S}^j $ is the symmetrization operator (a sum over all permutations on $\otimes^j {\mathbb C}^2$), given by
\begin{align}\label{eq:SymmOpDef}
{\mathbb S}^j =&\, \frac{1}{j!}\sum_{\sigma \in {\mathfrak S}_j} {\mathbb K}_\sigma\,,\\
{\mathbb S}^2 = &\,\frac{1}{2}\big( 1 + {\mathbb K}_{(12)}\big)\,,\quad 
{\mathbb S}^3 = \frac{1}{6}
\big( 1 + {\mathbb K}_{(123)}+ {\mathbb K}_{(132)}+ {\mathbb K}_{(12)}+ {\mathbb K}_{(23)}+ {\mathbb K}_{(31)}\big)\,.
\end{align}
Here, ${\mathbb K}_{(12)}$ is the elementary transposition operator (the switch operator) and 
${\mathbb K}_{(123)} = {\mathbb K}_{(23)} \circ {\mathbb K}_{(12)}$.
In the sequel, we will use the following relation
$$
 {\mathbb K}_{(12)}
= \oh \mbox{\boldmath{$\sigma$}}\cdot \mbox{\boldmath{$\sigma$}} + \oh \unit \otimes \unit\, .
$$
If we choose the $(2\times 2)$-elementary matrices  $\left\{ e_{pq} \right\}$, we have 
${\mathbb K}_{(12)} = \sum_{p,q=1}^2 e_{pq}\otimes e_{qp}$,
which in the above introduced notation reads
\[
{\mathbb K}_{(12)} = \left.\sum\right._{\sf x} {\sf x}\otimes  {\sf x'} \,.
\]
Using this transcription we have
\begin{equation}
\label{eq:basicbbbJjkDef}
2{\mathbb J}^{j,k} = 
 \left.\sum\right._{\sf x} {\sf x}^{\otimes_j}\otimes{\sf x'}^{\otimes_k}
- \textstyle{\frac 12}jk\unit^{\otimes_j}\otimes\unit^{\otimes_k}\, ,
\end{equation}
and finally, using \eqref{eq:stretchP}, we obtain the following. 
\bsz
\label{Proj-S-Calc}
In the above calculus the projectors \eqref{eq:stretchP} read\footnote{In the following formulae, the symmetrizers  $ {\mathbb S}^j$ 
are redundant, but for clarity of presentation are inserted here.
For notational convenience the reference representation (spin $\tfrac 12$\,, Dynkin labels 1)
has been shifted to the second tensor factor. }
\begin{align}
{\mathbb P}_{j\!+\!1}^{j,1} = &\, \frac{1}{j\!+\!1}\left.\sum\right._{\sf x} {\mathbb S}^j{\sf x}^{\otimes_j}{\mathbb S}^j \otimes {\sf x}' \!+\! \frac{1}{j\!+\!1}\unit^{j}\otimes \unit\,, \label{eq:ProjNeq2stretchDefn}\\
{\mathbb P}_{j-1}^{j,1} = &\, -\frac{1}{j\!+\!1}\left.\sum\right._{\sf x} {\mathbb S}^j{\sf x}^{\otimes_j} {\mathbb S}^j\otimes {\sf x}' \!+\! \frac{j}{j\!+\!1}\unit^{j}\otimes \unit\,; 
\label{eq:ProjNeq2squeezeDefn}
\end{align}
with inverses
\begin{align} 
\left.\sum\right._{\sf x} {\mathbb S}^j{\sf x}^{\otimes_j}{\mathbb S}^j \otimes {\sf x}' = &\,
j {\mathbb P}_{j\!+\!1}^{j,1} - {\mathbb P}_{j-1}^{j,1}\,, \qquad
\unit^{j}\otimes \unit \equiv  {\mathbb P}_{j\!+\!1}^{j,1} + {\mathbb P}_{j-1}^{j,1}\,.
\label{eq:ProjNeq2SweedlerInv}
\end{align}
\esz

\subsection{Evaluation of quasicharacters for $N=2\,, 3$ and $4$\,.}
\label{subsec:ExplicitNeq2n3characters}

To demonstrate the use of the projection formalism, we now show some illustrative steps leading to the evaluation of some $N=2$ and $N=3\,,4$ characters. To start with, let us consider  the following example.
\begin{exx}[The quasicharacter  $\chibig^4_{(31)}(u,v)$\,]
\mbox{}\\
\label{E-Quasich-431}
Noting that
$\mathbb{K}_{(132)}= \mathbb{K}_{(23)} \circ \mathbb{K}_{(12)}$, the first term of equation (\ref{eq:ProjNeq2stretchDefn}) will include the contribution
\[
\textstyle{\frac 16} \Tr\Big(\big((\left.\sum\right._{\sf x} 
{\sf x} \otimes \unit \otimes \unit +
\unit \otimes {\sf x} \otimes \unit +
\unit \otimes \unit \otimes  {\sf x})\otimes {\sf x'} \big) \cdot \big(({\sf y}u \otimes
{\sf z}{\sf y'} u \otimes {\sf z'}u)\otimes v \big)\Big)
\]
which entails three terms with triple sums over elementary matrices representing the pairwise switches. Using the following properties 
\begin{align*}
\left.\sum\right._{\sf x} \Tr({\sf x} a {\sf x'} b) = &\, \Tr(a) \Tr(b)\,, \qquad \left.\sum\right._{\sf x} \Tr({\sf x}a)\Tr({\sf x'}b) = \Tr(ab), \\
\left.\sum\right._{\sf x}\! \left.\sum\right._{\sf y} {\Tr({\sf xy}a)\Tr({\sf x'y'}  b) }=&\, 
{\Tr(a)\Tr(b)}\,,\qquad \left.\sum\right._{\sf x} \!\left.\sum\right._{\sf y} \Tr({\sf xy}a)\Tr({\sf y'x'}  b) = \Tr(\one)\Tr(ab)\,,
\end{align*}
with 
%
similar contributions from the remaining permutations contained in the symmetrizer, and using the Cayley-Hamilton identity 
$u^2 = \langle u\Tr(u)- \II$ for elements of $\SU(2)$, the  traces 
$\Tr(u^2)= \Tr(u)^2 - 2$ and $\Tr(\II)= 2$ yield the inhomogeneous
form\footnote{Reported as $\chibig_{(\fth\fh)2}(u,v)$ in table \ref{tab:Neq2exx}.}
\[
\chibig^4_{(31)}(u,v) = \,-  {\frac{1}{4}}\Tr(uv)- {\frac{1}{4}}\Tr(u)\Tr(v) + {\frac{3}{4}}\Tr(u)^2\Tr(uv)- {\frac{3}{8}}\Tr(u)^2\Tr(v)+ {\frac{1}{16}}\Tr(u)^3\Tr(v).
\]
\qeb
\end{exx}

For $N=3$ the relevant projections are developed pairwise according to the order of couplings.
As an example we consider the character with $\ulj=(j,1,1)$, diagonal in internal intermediate labels $k=j+1 = k'$:
\[
\chibig^{j+2}_{(j11),j\!+\!1,j\!+\!1}(u,v,w) = \Tr\Big({\mathbb P}^{1,j\!+\!1}_{j\!+\!2}\!\cdot\! \big( {\mathbb P}^{1,j}_{j\!+\!1} \!\cdot\! \pi_j(u)\otimes v\big) \otimes w \Big)
= \Tr\Big({\mathbb P}^{1,j\!+\!1}_{j\!+\!2}\!\cdot\! \big( {\mathbb P}^{1,j}_{j\!+\!1} \otimes \one
\!\cdot\! \pi_j(u)\otimes v \otimes w \big)\Big)\,.
\]

As a final example, we consider the extension of the method to the $N=4$ quasicharacter
$\chibig^{j\!+\!3\,j\!+\!2}_{(j111),j\!+\!1\,j\!+\!2,j\!+\!1\,j\!+\!2}$\,.
This requires a composition of three projectors,
\[
{\mathbb P}^{1,j\!+\!2}_{j\!+\!3}\!\cdot\!( {\mathbb P}^{1,j\!+\!1}_{j\!+\!2}\!\otimes 
\II\! )\!\cdot\! ( {\mathbb P}^{1,j}_{j\!+\!1} \!\otimes\! \II\!\otimes\! \II)\,,
\]
whose switch operator form will entail symmetrizations
of the form $ {\mathbb S}^{j\!+\!2} \otimes \II $ composed with
$ {\mathbb S}^{j\!+\!1} \otimes \II \otimes \II$ on the appropriate subspaces.

\begin{exx}[The quasicharacter $ \chibig^4_{(1111)2\,3,2\,3}(r,s,t,u)$\,]
\mbox{}\\
\noindent
\label{eq:Quasi111123234}
For the case $j=1$, the switch operator form is\footnote{Using $\cdot $ as shorthand for tensor product.}
\begin{align*}
&\, 24\chibig^4_{(1111)2\,3,2\,3}(r,s,t,u)=\\
&\, 
\quad \sum \Tr \left\{\rule{0pt}{10pt}\big(\sz \cdt 1 \cdt 1 \cdt \sz' +  1 \cdt \sz \cdt 1 \cdt \sz'+1 \cdt 1 \cdt \sz \cdt \sz') +1\cdt 1\cdt  1\cdt 1\big)\cdot 
\big(\tfrac 23(\sy \cdt 1 \cdt \sy' \cdt 1 + 1 \cdt \sy \cdt \sy' \cdt 1
+ \sy \cdt \sy' \cdt 1 \cdt 1) +1\cdt 1\cdt  1\cdt 1\big)\cdot\right.\\
&\,\qquad \left. \rule{0pt}{10pt}\cdot \big(\tfrac 13(\sx \cdt \sx' \cdt 1 \cdt 1+
 \sx' \cdt 1 \cdt \sx\cdt 1+1 \cdt \sx \cdt \sx'\cdt 1)+1\cdt 1\cdt  1\cdt 1\big)\cdot (r\cdt s\cdt t\cdt u)\,\right\}\,.
\end{align*}
Here the overall symmetrization 
\[
({\mathbb S}^{3} \otimes \II)\cdot ({\mathbb S}^{2} \otimes \II \otimes \II)
\equiv ({\mathbb S}^{3} \otimes \II) 
\]
has been implemented to replace contributions from $\bJ^{1,1}$ and 
$\bJ^{2,1}$ (Proposition \ref{Proj-S-Calc} above) with the appropriately re-weighted forms, for example 
\[
\sx \cdt \sx' \cdt 1 \cdt 1 \Rightarrow \tfrac 13 \big(\sx \cdt \sx' \cdt 1 \cdt 1+
 \sx' \cdt 1 \cdt \sx\cdt 1+1 \cdt \sx \cdt \sx'\cdt 1\big)\,.
\]
The computation entails accumulating all the contributions from the 
64 terms and attributing them to trace types with the appropriate strengths. An example is (with overall trace understood):
\begin{align*}
\sz r \cdot \sy \sx s \cdot \sy' \sx' t \cdot \sz' u \Rightarrow&\,
\Tr ( r u )  \Tr ( \sy \sx s \cdot \sy' \sx' t  )
\Rightarrow \Tr ( r u ) \Tr ( \sx s \sx' t  )
\Rightarrow \Tr ( r u ) \Tr ( s )\Tr ( t  )\,.
\end{align*}
The final result is\footnote{Reported as $\chibig^2_{(\fth\fh\fth\fh) 1 \tfrac 32, 
1 \tfrac 32}(r,s,t,u)$ in table \ref{tab:Neq3exx}.}
\begin{align*}
24\chibig_{(1111) 23,23}^4 =&\,
\big(\frac 23 \Tr( rstu +rsut+ rust) + \frac 89
\Tr( rtsu+ rtus + ruts)
\big) \\
&\, + 
\big(\Tr( s )\Tr( rtu+rut) +\Tr( r )\Tr( stu+sut) 
\Tr( t )\Tr( rsu+rus)  +\frac 23 \Tr( u )\Tr( rst+rts) \big)\\
&\, +\big(\Tr( rs ) \Tr( t) \Tr( u )
+\frac{13}{9}(\Tr( ru ) \Tr( s) \Tr( t )+\Tr( su ) \Tr( r) \Tr( t )) + 
\frac 49 \Tr( ut ) \Tr( r) \Tr( s )\big)\\
&\, + 2\big(\Tr( rs ) \Tr( ut) + \Tr( ru ) \Tr( st) +
\Tr( rt ) \Tr( su)\big)
+\frac{13}{9}\Tr( r ) \Tr( s) \Tr( t ) \Tr( u )\,.
\end{align*}
\qeb
\end{exx}

In tables \ref{tab:Neq2exx} and \ref{tab:Neq3exx} we give further explicit values of some irreducible $\chibig$'s, for $N=2$ and $N=3, N=4$\,, respectively, computed via projection techniques, as illustrated in the above examples.
\subsection{Products of $N=2$ quasicharacters and $9j$ symbols.}
\label{subsec:ExplicitNeq2products}
%
%

It turns out that, for $N=2$, one can explicitly derive the general multiplication law for quasicharacters via a corresponding manipulation of the tensor products of the characteristic projectors themselves. We will see that, given this law, the general multiplication law for the quasicharacters given by Proposition \ref{Gen-MultLaw} and Lemma \ref{L-symbols} follows immediately. We have
\begin{align*}
    \bP_{j_1j_2}^k \otimes\bP_{j'_1j'_2}^{k'} =&\,
    \sum | k m, j_1j_2\rangle \langle k m, j_1j_2 |\otimes | k' m', j'_1j'_2\rangle \langle k' m', j'_1j'_2 |\,,
\end{align*}
leading, after uncoupling of the paired angular momenta, to
\begin{align*}
    \bP_{j_1j_2}^k \otimes\bP_{j'_1j'_2}^{k'} =&\, \sum |j_1 m_1, j_2 m_2, j'_1 m'_1, j'_2 m'_2\rangle \langle j_1 m_1, j_2 m_2  | k m, j_1j_2 \rangle
    \langle j'_1 m'_1, j'_2 m'_2  | k' m', j'_1j'_2 \rangle\mbox{\boldmath{$\cdot$}}\\
    &\,\hskip1cm\mbox{\boldmath{$\cdot$}}\langle k m, j_1j_2  |j_1 n_1, j_2 n_2 \rangle \langle k' m', j'_1j'_2 |j'_1 n'_1 ,j'_2 n'_2 \rangle
    \langle j_1 n_1, j_2 n_2, j'_1 n'_1, j'_2 n'_2 |\,.
\end{align*}
In the following manipulations we assume summation over repeated indices, including coupled angular momenta. In the last step, the summations
\[
(\cdots)_{m,m'} (\cdots)_{m,m'} \equiv 
(\cdots)_{m,m'} (\cdots)_{m^*,m'{}^*} \delta_{m,m^*}\delta_{m',m'{}^*} 
\]
are re-written as
\[
(\cdots)_{m,m'} (\cdots)_{m^*,m'{}^*} 
C^{k k' K^*}_{m m^* M^*}C^{k k' K^*}_{m' m'{}^* M^*}\,,
\]
with the $\delta$-constraint being implemented by the unitarity of the corresponding Clebsch-Gordan coefficient matrix (summation over $K^*,M^*$ with $K^* \in \langle k,k'\rangle$). We have then
\begin{align*}
    &\,  |j_1 m_1, j_2 m_2, j'_1 m'_1, j'_2 m'_2\rangle 
    C^{j_1j_2k}_{m_1m_2m}C^{j'_1j'_2k'}_{m'_1m'_2m'}C^{j_1j_2k}_{n_1n_2n}C^{j'_1j'_2k'}_{n'_1n'_2n'}
    \langle j_1 n_1, j_2 n_2, j'_1 n'_1, j'_2 n'_2 |\\
   =&\, |J_1M_1,J_2M_2\rangle \langle J_1M_1, j_1j'_1|j_1 m_1 ,j'_1 m'_1\rangle
    \langle J_2M_2, j_2j'_2|j_2 m_2, j'_2 n'_2\rangle \mbox{\boldmath{$\cdot$}} \\
    &\,\hskip1cm\mbox{\boldmath{$\cdot$}} C^{j_1j_2k}_{m_1m_2m}C^{j'_1j'_2k'}_{m'_1m'_2m'}C^{j_1j_2k}_{n_1n_2n}C^{j'_1j'_2k'}_{n'_1n'_2n'}
    \langle j_1n_1, j'_1n'_1|J_1M'_1,j_1j'_1\rangle \langle j_2n_2, j'_2n'_2|J_2M'_2,j_2j'_2\rangle
\langle J_1M'_1,J_1M'_2|\\
    =&\, |J_1M_1,J_2M_2\rangle C^{j_1j'_1 J_1}_{m_1m'_1M_1}C^{j_2j'_2 J_2}_{m_2m'_2M_2}\mbox{\boldmath{$\cdot$}}
    C^{j_1j_2k}_{m_1m_2m}C^{j'_1j'_2k'}_{m'_1m'_2m'}C^{j_1j_2k}_{n_1n_2m}C^{j'_1j'_2k'}_{n'_1n'_2m'}\mbox{\boldmath{$\cdot$}}
    C^{j_1j'_1 J_1}_{n_1n'_1M'_1}C^{j_2j'_2 J_2}_{n_2n'_2M'_2}\langle J_1M'_1,J_1M'_2|\,\\
    =&\, |KM\rangle C^{J_1J_2 K}_{M_1M_2M}C^{j_1j'_1 J_1}_{m_1m'_1M_1}C^{j_2j'_2 J_2}_{m_2m'_2M_2}\mbox{\boldmath{$\cdot$}} C^{j_1j_2k}_{m_1m_2m}C^{j'_1j'_2k'}_{m'_1m'_2m'}C^{j_1j_2k}_{n_1n_2m}C^{j'_1j'_2k'}_{n'_1n'_2m'}\mbox{\boldmath{$\cdot$}}  C^{j_1j'_1 J_1}_{n_1n'_1M'_1}C^{j_2j'_2 J_2}_{n_2n'_2M'_2}C^{J_1J_2K'}_{M'_1M'_2M'}\langle K'M'|\,\\
    \equiv &\,|KM\rangle C^{J_1J_2 K}_{M_1M_2M}C^{j_1j'_1 J_1}_{m_1m'_1M_1}C^{j_2j'_2 J_2}_{m_2m'_2M_2}\mbox{\boldmath{$\cdot$}} C^{j_1j_2k}_{m_1m_2m}C^{j'_1j'_2k'}_{m'_1m'_2m'^*}C^{j_1j_2k}_{n_1n_2m^*}C^{j'_1j'_2k'}_{n'_1n'_2m'{}^*}
    \mbox{\boldmath{$\cdot$}} \\
&\,\hskip1cm \mbox{\boldmath{$\cdot$}}\delta_{mm^*}\delta_{m'm'{}^*} \mbox{\boldmath{$\cdot$}}C^{j_1j'_1 J_1}_{n_1n'_1M'_1}C^{j_2j'_2 J_2}_{n_2n'_2M'_2}C^{J_1J_2K'}_{M'_1M'_2M'}\langle K'M'|\,\\
    =&\,|KM\rangle C^{J_1J_2 K}_{M_1M_2M}C^{j_1j'_1 J_1}_{m_1m'_1M_1}C^{j_2j'_2 J_2}_{m_2m'_2M_2}\mbox{\boldmath{$\cdot$}} 
    C^{j_1j_2k}_{m_1m_2m}C^{j'_1j'_2k'}_{m'_1m'_2m'^*}C^{j_1j_2k}_{n_1n_2m^*}C^{j'_1j'_2k'}_{n'_1n'_2m'{}^*}\mbox{\boldmath{$\cdot$}}\\
&\,
    \mbox{\boldmath{$\cdot$}} C^{k k' K^*}_{m m' M^*}C^{k k' K^*}_{m^* m'{}^* M^*} \mbox{\boldmath{$\cdot$}} 
    C^{j_1j'_1 J_1}_{n_1n'_1M'_1}C^{j_2j'_2 J_2}_{n_2n'_2M'_2}C^{J_1J_2K'}_{M'_1M'_2M'}\langle K'M'|\,.
\end{align*}
Finally regrouping the 12 Clebsch-Gordan factors we have
\begin{align}
\bP_{j_1j_2}^k \otimes\bP_{j'_1j'_2}^{k'}  =&\,|KM,J_1J_2\rangle \langle K'M',J_1J_2|\mbox{\boldmath{$\cdot$}} \nonumber \\
&\,\left[ C^{j_1j_2k}_{m_1m_2m}C^{j'_1j'_2k'}_{m'_1m'_2m'}C^{k k' K^*}_{m m' M^*}\mbox{\boldmath{$\cdot$}} 
C^{j_1j'_1 J_1}_{m_1m'_1M_1}C^{j_2j'_2 J_2}_{m_2m'_2M_2}C^{J_1J_2 K}_{M_1M_2M}\right] \mbox{\boldmath{$\cdot$}}\\
&\,\left[ C^{j_1j_2k}_{n_1n_2m^*}C^{j'_1j'_2k'}_{n'_1n'_2m'{}^*} C^{k k' K^*}_{m^* m'{}^* M^*}
\mbox{\boldmath{$\cdot$}} C^{j_1j'_1 J_1}_{n_1n'_1M'_1}C^{j_2j'_2 J_2}_{n_2n'_2M'_2}C^{J_1J_2K'}_{M'_1M'_2M'}\right]\,. 
\label{eq:9jImplicit}
\end{align}
Each bracket contains independent summations (over magnetic states $m_1,m_2, m'_1,m'_2$
and $n_1,n_2, n'_1,n'_2$ respectively) and represents the overlap of totally coupled 4 spin states
on 2 different 4 leaf trees. This implements 
$\delta_{K^*K}\delta_{K^*K'}\delta_{M^*M}\delta_{M^*M'}$ and identifies the brackets as the 
square of the corresponding $9j$ symbol, independent of label $M$\,, and the summation 
becomes as before
\footnote{Following the definition in \cite{biedenharn1981angular}.}
\begin{align}  
\label{eq:ProjProd} 
\bP_{j_1j_2}^k \otimes \bP_{j'_1j'_2}^{k'} =&\, 
 \sum_{J_1, J_2, K} [J_1][J_2] [k][k']\left.\left\{\begin{array}{ccc} j_1 & j_2 & k\\j'_1 & j'_2 & k' \\
J_1 & J_2 & K \end{array}\right\}\right.^{\!\!2} \bP^{K}_{J_1J_2} .
\end{align}
Note that each grand projector $\bP^{K}_{J_1J_2}$ is in fact a projection
over $H_{J_1}\otimes H_{J_2}$ \emph{which is a $[J_1]\cdot[J_2]$-dimensional subspace of the full 
four-fold factor space $H_{j_1}\otimes H_{j_2}\otimes H_{j'_1}\otimes H_{j'_2}$}\,, that is, the
spaces $H_{J_1}\otimes H_{J_2}$ are isomorphic copies of the corresponding spin irreps here realized in the total space. 
The formula for pointwise product of quasicharacters now
follows directly from that for the (tensor) product of the associated projectors:
\begin{align*}
  \chibig^{j_1, j_2}_{J_1}(u,v)\mbox{\boldmath{$\cdot$}} 
  \chibig^{j'_1, j'_2}_{J_2}(u,v) =&\,
\Tr \big( \bP_{j_1j_2}^{J_1} \mbox{\boldmath{$\cdot$}} D_{j_1}(u) \otimes D_{j_2}(v) \big)\mbox{\boldmath{$\cdot$}}
\Tr \big( \bP_{j'_1j'_2}^{J_2}\mbox{\boldmath{$\cdot$}} D_{j'_1}(u) \otimes D_{j_2}(v) \big)\\
=&\,
\Tr\Big(\big(\bP_{j_1j_2}^{J_1}\otimes\bP_{j'_1j'_2}^{J_2}\big) \mbox{\boldmath{$\cdot$}}
\big(D_{j_1}(u) \otimes D_{j_2}(v) \otimes D_{j'_1}(u) \otimes D_{j'_2}(v)\big)\Big).
\end{align*}
Thus finally, we obtain the following. 
\begin{lem}[Pointwise product of $N=2$ irreducible quasicharacters]
\label{lem:PtWiseNeq2}
\begin{align}
\label{eq:PtwiseNeq2Full}
\chibig_{j_1, j_2}^{k}(u,v)\mbox{\boldmath{$\cdot$}}
  \chibig_{j'_1, j'_2}^{k'}(u,v) =&\,
\sum_{J_1, J_2, K} [J_1][J_2] [k][k']\left.\left\{\begin{array}{ccc} j_1 & j_2 & k\\j'_1 & j'_2 & k' \\
J_1 & J_2 & K \end{array}\right\}\right.^{\!\!2} \chibig_{J_1, J_2}^{K}(u,v)\,.
\end{align}
\bx
\end{lem}

Table \ref{tab:9jcalcs} lists the derived $9j$ symbols (after normalization) for cases evaluated explicitly in section \ref{subsec:ExplicitNeq2products} (with standard 
angular momentum labelling).

Compiling the set of structure constants for pointwise multiplication of quasicharacters would of course be immediate if the entire character table were at hand. 
As a further application of the projection operator method of construction, 
we treat the computation of certain $N=2$ products directly, exploiting the 
formalism described above (sections \ref{sec:SpinChainsProjectors},\ref{subsec:ExplicitNeq2n3characters}), to develop explicit product formulae for 
some special cases (multiplication by fundamental group characters).
As is known \cite{Fuchs2018costratificationMR,jarvis2021quasicharactersMR} from the formal viewpoint (section \ref{sec:Introduction} above,
and sections \ref{subsec:Properties}, \ref{subsec:ExplicitNeq2Products9j} below), the multiplicative structure constants are (squares of) Racah $9j$ coefficients, so that the present computations amount to \textit{de novo} computation of some special $9j$ Racah symbols, and serve to verify the overall formalism.

To explain our method, in the sequel, we are going to establish the multiplication rules for the following special types:
$$
\Tr(u)\cdot \chibig_{j,1}^\ell(u,v), \quad \Tr(v)\cdot\chibig_{j,1}^\ell(u,v)\,, \quad \ell=j \pm 1 \, ,
$$
for arbitrary $j$. This amounts to evaluating pointwise products, of some instances of $\chibig$ constructed formally above  in terms of the switch operator calculus. These basic multiplicands of course happen to be standard irreducible, $\SU(2)$ spin-$\oh$ characters:
$\Tr(u)\equiv \chibig_{10}^1(u,v)$\,, $\Tr(v)\equiv \chibig_{01}^{1}(u,v)$\,.
We proceed with the calculation for the maximal $\ell$ cases $\ell=j\!+\! 1$\,; the expansions for $\ell=j\!-\! 1$ can be inferred using similar methods. 

As a preliminary remark, note that from equation (\ref{G-MF-1}), the quasicharacter product necessitates decoupling the constituent paired angular momenta, followed by subsequent rearrangement and recoupling. In the case of 
the pointwise product with $\Tr(u)$\,, this latter step will entrain the tensor product
reduction $H_{1} \otimes H_j \cong H_{j+1} \oplus H_{j-1}$ within the
$(j+1)$-fold tensor power $\otimes^{j+1}{\mathbb C}^2$\,. This is consistent with the
standard diagrammatic Littlewood-Richardson rule
\[
\scalebox{.6}{$\young(~)$}\times
\scalebox{.6}{$\young(~~~)$}\,\cdots
\scalebox{.6}{$\young(~)$}\,\cong\,
\scalebox{.6}{$\young(~~~)$}\,\cdots
\scalebox{.6}{$\young(~~)$}\, +\,
\scalebox{.6}{\raisebox{-0.15cm}{$\young(~~~,~)$}}\raisebox{.21cm}{$\,\cdots$}
\scalebox{.6}{\raisebox{0.28cm}{$\young(~)$}}\,,
\]
which corresponds to the sum of the totally symmetric plus mixed symmetry Young projectors $\II^{j+1} = {\mathbb S}^{j+1}+{\mathbb M}^{j+1}$\,. Noting the definition
(\ref{eq:SymmOpDef}), these projectors (and the corresponding resolution of the identity) are given recursively as
\begin{align}
{\mathbb S}^{j\!+\!1}=&\,
\frac{1}{(j\!+\!1)}\big({\mathbb I} + {\mathbb K}_{(j\!+\!1 1)} +{\mathbb K}_{(j\!+\!1 2)} +\cdots + {\mathbb K}_{(j\!+\!1 j)}\big)\big({\mathbb S}^j\otimes\one\big)\,,\label{eq:SymmDef}\\
{\mathbb M}^{j\!+\!1}=&\,
\frac{1}{(j\!+\!1)}\big(j {\mathbb I} - {\mathbb K}_{(j\!+\!1 1)} -{\mathbb K}_{(j\!+\!1 2)} -\cdots - {\mathbb K}_{(j\!+\!1 j)}\big)\big({\mathbb S}^j\otimes\one\big)\,.\label{eq:MixedDef}
\end{align}
In the case of the pointwise product with $\Tr(v)$, the corresponding 
tensor product is simply $H_{1} \otimes H_1 \cong H_{2} \oplus H_{0}$\,,
and the mixed symmetry projector coincides with the antisymmetrizer,
$\II^{2} = {\mathbb S}^{2}+{\mathbb A}^{2}$\,.\\

\noindent
\textbf{Evaluation of $\Tr(u)\cdot \chibig_{j,1}^{j\!+\! 1}(u,v)$:}\\
Recall the explicit projection operator (\ref{eq:stretchP}),
and the switch operator equivalent\footnote{Or (\ref{eq:stretchP}), (\ref{eq:ProjNeq2squeezeDefn}), respectively, for the cases $\ell = j\!-\! 1$\,.},
(\ref{eq:ProjNeq2stretchDefn}). 
As above, we note the isomorphic description of spin-$(\hj\!-\!\oh)$ (Dynkin label $(j\!-\!1)$) recovered from 
the $(j\!+\!1)$-fold tensor product of the fundamental representation (rather than the $(j-1)$-fold tensor power) using the mixed symmetry projection,
\begin{equation}
\widetilde{\pi}_{j-1}(a) := {\mathbb M}^{j+1} u^{\otimes_{j+1}}  {\mathbb M}^{j+1}\,.
\label{eq:TaujDefn}
\end{equation}
The invariant ${\mathbb J}^{j-1,1}$ must satisfy the appropriate characteristic equation, but when rewritten in terms of switch operator insertions, equation (\ref{eq:basicbbbJjkDef}) leads to a different coefficient of the identity operator. Using equations (\ref{eq:TaujDefn}), 
(\ref{eq:ProjNeq2stretchDefn}), and (\ref{eq:ProjNeq2squeezeDefn}) for the case in hand
(tensoring spin-$(\hj -\oh)$ (but working with its realization in $\otimes^{j+1}{\mathbb C}^2$)\, with the fundamental, we have for the corresponding projection operators
\begin{align}
\label{eq:PjprojectorsJtoM}
{\mathbb P}_{j}^{(j\!-\!1,1)} = &\,  \frac{2\widetilde{\mathbb J}^{(j-1,1)} + (\hj+\textstyle{\frac 12})}{j}\, ,\quad 
{\mathbb P}_{j\!-\!2}^{(j\!-\!1,1)} = \frac{-2\widetilde{\mathbb J}^{(j-1,1)} +(\hj-\textstyle{\frac 12})}{j}\,, \nonumber\\
\mbox{whence} \quad \widetilde{\mathbb P}^{j\!-\!1, 1}_j = &\, \frac{1}{j}\left.\sum\right._{\sf x} {\mathbb M}^{j+1}{\sf x}^{\otimes_{j+1}}{\mathbb M}^{j+1} \otimes {\sf x}' \,,\\
\mbox{and} \quad\widetilde{\mathbb P}^{j\!-\!1, 1}_{j\!-\!2} = &\, -\frac{1}{j}\left.\sum\right._{\sf x}{\mathbb M}^{j+1} {\sf x}^{\otimes_{j+1}}{\mathbb M} \otimes {\sf x}' + \one^{j+1}\otimes \one\,.\nonumber
\end{align}
For the product $\Tr(u)\cdot \chibig_{j,1}^{j\!+\!1}(u,v)$ we have
from equation (\ref{eq:ProjNeq2stretchDefn}), 
\begin{align}
\Tr(u)\chibig_{j,1}^{j+1}= & \,\frac{1}{j+1} \left.\sum\right._{\sf x} \Tr\big( ({\sf x}^{\otimes_j}\cdot u^{\otimes_j} \cdot{\mathbb S}^j)\otimes u\big)\Tr({\sf x}'v) + 
\frac{1}{j+1}\Tr\big((u^{\otimes_j} \cdot{\mathbb S}^j)\otimes u\big)\Tr(v)\,
\label{eq:TrUfirstform}
\\
\equiv &\, \frac{1}{j+1}{[}A{]} + \frac{1}{j+1}{[}B{]}\, . \nonumber
\end{align}
Term ${[}B{]}$ entails a product
of fundamental characters, and by the composition rule for angular momenta we have
\begin{align}
{[}B{]}=\Tr\big(\pi_j(u)\big) Tr(u)Tr(v) = &\,
Tr\big(\pi_{j\!+\!1}(u)\big)Tr(v) + Tr\big(\pi_{j\!-\!1}(u)\big)Tr(v)\,.
\label{eq:TrUfirstformCharProd}
\end{align}
Using (\ref{eq:SymmDef}), (\ref{eq:MixedDef})\,, term ${[}A{]}$ is on the other hand
\begin{align}
{[}A{]} =&\,\Tr\big( ({\sf x}^{\otimes_j}\cdot u^{\otimes_j} \cdot{\mathbb S}^j)\otimes u\big)
\Tr({\sf x}'v)
= \Tr\big( ({\sf x}^{\otimes_j}\otimes \one)\cdot u^{\otimes_{j\!+\!1}} \cdot({\mathbb S}^j\otimes \one)\big)\Tr({\sf x}'v) \nonumber \\
= &\,\Tr\big( ({\sf x}^{\otimes_j}\otimes \one)\cdot u^{\otimes_{j\!+\!1}} \cdot{\mathbb S}^{j\!+\!1}\big)\Tr({\sf x}'v)
+ \Tr\big( ({\sf x}^{\otimes_j}\otimes \one)\cdot u^{\otimes_{j\!+\!1}} \cdot{\mathbb M}^{j\!+\!1}\big)\Tr({\sf x}'v) 
\label{eq:SplitTrUfirstform} \\
\equiv &\, {[}C{]} + {[}D{]}\,.
\end{align}
Using the cyclicity of the trace, the coproduct $\sf x^j \otimes \II$ in {[}C{]} may be
smeared across the $(j\!+\!1)$-fold tensor product, so that
\begin{align}
{[}C{]}= &\, \frac{j}{(j\!+\!1)}
\Tr\big( {\sf x}^{\otimes_{j\!+\!1}}\cdot u^{\otimes_{j\!+\!1}} \cdot{\mathbb S}^{j\!+\!1}\big)\Tr({\sf x}'v)\,.
\label{eq:SplitTrUfirstform1Final}
\end{align}
Proceeding with the second, mixed term, we may use (from Young diagram 
\scalebox{.6}{\raisebox{-0.15cm}{$\young(~~~,~)$}}\raisebox{.21cm}{$\,\cdots\!$}
\scalebox{.6}{\raisebox{0.28cm}{$\young(~)$}}\,)
\[
{\mathbb M}^{j\!+\!1}=
\frac{j}{(j\!+\!1)}\big( {\mathbb I} - {\mathbb K}_{(j\!+\!1 1)}\big)\big({\mathbb S}^j\otimes\one\big)\,,
\]
as proxy for (\ref{eq:MixedDef}), in the presence of the symmetrization over positions $1,2,\cdots, j$\, implemented by $\big({\mathbb S}^j\otimes\one\big)$ across the trace: 
\begin{align}
\label{eq:Mpreparation}
{[}D{]}
=&\, \frac{j}{j\!+\!1} \Tr\big( ({\sf x}^{\otimes_j}\otimes \one)\cdot u^{\otimes_{j\!+\!1}} \cdot({\mathbb S}^j\otimes \one)\big)\Tr({\sf x}'v) -\\ \nonumber
&\, - 
\frac{j}{j\!+\!1}\left.\sum\right._{\sf y}\Tr\big( ({\sf x}^{\otimes_j}\otimes \one)\cdot u^{\otimes_{j\!+\!1}}
\cdot({\sf y}\otimes \one \cdots \otimes {\sf y'}) \cdot({\mathbb S}^j\otimes \one)\big)\Tr({\sf x}'v) \\
= & \, \frac{j}{j\!+\!1}\left.\sum\right._{\sf y}\Tr\big( ({\sf x}^{\otimes_j})\cdot u^{\otimes_{j}}
 \cdot({\mathbb S}^j)\big)\Tr(u)\Tr({\sf x}'v)\\  \nonumber
&\,
- \frac{j}{j\!+\!1}\left.\sum\right._{\sf x, y}\Tr\big( ({\sf x}^{\otimes_j})\cdot (u{\sf y}\otimes u^{\otimes_{j-1}})
\cdot {\mathbb S}^j\big)\Tr(u{\sf y}')\Tr({\sf x}'v)\,. 
\end{align}
Now note that $\left.\sum\right._{\sf y} \Tr(\cdots u {\sf y} \cdots )\Tr(u{\sf y}') \equiv \Tr(\cdots u^2 \cdots)$
which can be replaced by $u^2 = u \Tr(u) -\one$\,; thus the first two contributions
cancel leaving
\[
{[}D{]}= \frac{j}{j\!+\!1}Tr\big( ({\sf x}^{\otimes_j})\cdot (\one \otimes u^{\otimes_{j-1}})
\cdot {\mathbb S}^j\big)Tr({\sf x}'v)\,.
\]
By similar techniques, writing now ${\sf x}^{\otimes_j} = {\sf x}\otimes \one^{j-1} + \one \otimes {\sf x}^{\otimes_{j-1}}$, and implementing (\ref{eq:SymmDef}) for ${\mathbb S}^j$\,, further consideration of tensor product structure across the trace leads to 
\begin{align}
\label{eq:SplitTrUfirstform2Final}
{[}D{]}= &\,\frac{1}{j\!+\!1}\Tr(u^{\otimes_{j-1}}{\mathbb S}^{j-1}) \Tr(v) 
+ \frac{j\!+\!2}{j\!+\!1} \left.\sum\right._{\sf x} \Tr({\sf x}^{\otimes_{j-1}}\cdot u^{\otimes_{j-1}}\cdot {\mathbb S}^{j-1}) \Tr({\sf x}'v) \,.
\end{align}
With equations (\ref{eq:SplitTrUfirstform1Final}), (\ref{eq:SplitTrUfirstform2Final})
for the first and second terms of ${[}A{]}$\,,
together with the switch independent part ${[}B{]}$\,,
we are in a position to rewrite contributions to (\ref{eq:TrUfirstform}) as combinations of projection operators, 
using equation (\ref{eq:ProjNeq2SweedlerInv})\,, and hence to infer the expansion of the pointwise product in question. Collecting terms we arrive at the following final form.

\begin{lem}[Pointwise product $\chibig_{1,0}^{1}(u,v)\chibig_{j,1}^{j\!+\!1}(u,v)$]
\label{lem:TruProdForm}
\begin{align}
\label{eq:TruProdForm}
\chibig_{1,0}^{1}(u,v)\chibig_{j,1}^{j\!+\!1}(u,v)=&\, \chibig_{j\!+\!1,1}^{j\!+\!2}(u,v) + \frac{1}{(j\!+\!1)^2}\chibig_{j\!+\!1,1}^{j}(u,v) + \frac{j(j\!+\!2)}{(j\!+\!1)^2}\chibig_{j\!-\!1,1}^{j}(u,v) + 0\cdot \chibig_{j\!-\!1,1}^{j\!-\!2}(u,v) .
\end{align}
\end{lem}
\mbox{}\hfill $\Box$\\
\noindent
\textbf{Evaluation of $Tr(v)\cdot \chibig_{j,1;j\!+\! 1}(u,v)$:}\\
Turning to the corresponding calculation of the pointwise product with $Tr(v)$, we have
\begin{align*}
\Tr(v)\chibig_{j,1}^{j+1}(u,v) = &\, \Tr(v)\Tr\big({\mathbb P}_{j+1}^{(j,1)} \cdot \pi_j(u)\otimes v\big) \\
=&\, \frac{1}{j+1}\Tr\Big(2{\mathbb J}^{j,1}\otimes \one \cdot \big(u^{\otimes_j} {\mathbb S}^j \otimes v\big)\otimes v \Big)
+ \frac{\hj+1}{j+1}\Tr\Big(\one^{j+1} \otimes \one \cdot \big(\big(u^{\otimes_j} {\mathbb S}^j\otimes v\big)\otimes v \Big) 
\end{align*}
which becomes in terms of the correlated sums (see (\ref{eq:basicbbbJjkDef}) above) and projection operators 
${\mathbb I}^2 = {\mathbb S}^2+{\mathbb A}^2$\,,
\begin{align*}
\Tr(v)\chibig_{j,1}^{j+1}(u,v) = &\, \frac{1}{j+1}\left.\sum\right._{\sf x} \Tr\Big(\big({\sf x}^{\otimes_j} \otimes \textstyle{\frac 12}\big({\sf x'}^{\otimes_2}\big)\! \cdot\!\big(u^{\otimes_j} {\mathbb S}^j\big) \otimes {\mathbb S}^2\big(v\otimes v \big)\Big)+
 \\
 &\,+ \frac{1}{j+1}\left.\sum\right._{\sf x} \Tr\Big(\big({\sf x}^{\otimes_j}\otimes\textstyle{\frac 12}\big({\sf x'}^{\otimes_2}\big)\big)\! \cdot\! \big(\big(u^{\otimes_j} {\mathbb S}^j\big) \otimes {\mathbb A}^2\big(v\otimes v \big)\Big)\\
&\, + \frac{1}{j+1}
\Tr\Big( \big(\big(u^{\otimes_j} {\mathbb S}^j\big)
\otimes {\mathbb S}^2\big(v\otimes v \big)\Big)+ 
 \frac{1}{j+1}\Tr\Big(\big(u^{\otimes_j} {\mathbb S}^j\big)
\otimes {\mathbb A}^2\big(v\otimes v \big)\Big)\,, 
\end{align*}
where the insertion ${\sf x}' \otimes \one$ has been averaged as $\oh {\sf x'}^{\otimes_2}$ in the presence of overall (anti)symmetrization.  Thus 
\begin{align}
\Tr(v)\chibig_{j,1}^{j+1} = &\, \frac{1}{j+1}\sum \Tr\Big(\big({\mathbb J}^{j,2}+ (\hj+1)\big)\! \cdot\! \big(u^{\otimes_j} {\mathbb S}^j\big) \otimes {\mathbb S}^2\big(v\otimes v \big)\Big)+
\nonumber  \\
 &\,+ \frac{(\hj+1)}{j+1}\sum \Tr\Big(\big(u^{\otimes_j} {\mathbb S}^j\big) \otimes {\mathbb A}^2\big(v\otimes v \big)\Big)\,,
\label{eq:ReorganiseProdTrb} 
\end{align}
again using equation ( (\ref{eq:basicbbbJjkDef}) and recognising that
$\big({\mathbb J}^{j,2}\cdot \big(u^{\otimes_j} {\mathbb S}^j\big)
\cdot \big(v^{\otimes_2} {\mathbb A}^2\big)$ vanishes as $\big(v^{\otimes_2} {\mathbb A}^2\big)$ is one-dimensional. Introducing the projection operators
${\mathbb P}_{\ell}^{(j,2)} $\,, $\ell =j,  j\!\pm\!2 $ (Lemma \ref{lem:SU2Projs}, equations (\ref{eq:Projectorsj2})), 
we reorganize the contributions to (\ref{eq:ReorganiseProdTrb}) above leading 
to the final form\footnote{The expansion is determined up to overall proportionality
by the absence in (\ref{eq:ReorganiseProdTrb}) of $\big({\mathbb J}^{j,2}\big){}^2$\,.}

\begin{lem}[Pointwise product $\chibig_{0,1}^{1}(u,v)\chibig_{j,1}^{ j\!+\!1}(u,v)$]
\label{lem:TrvProdForm}
\begin{equation}
\label{eq:TrvProdForm}
\chibig_{0,1}^{1}(u,v)\, \chibig_{j,1}^{j+1}(u,v) = {\chibig_{j,2}^{j+2}(u,v) + \frac{\hj}{j+1}\chibig_{j,2}^{j}(u,v) + \frac{\hj+1}{j+1}\chibig_{j,0}^{j}}(u,v).
\end{equation}
\end{lem}
\mbox{}\hfill $\Box$\\[-.5cm]
{\begin{rem}[\emph{Ab initio} $9j$ Racah coefficients]\mbox{}\\
In the expansions (\ref{eq:TruProdForm}), (\ref{eq:TrvProdForm}) resulting from
the pointwise product computations  
the coefficients arising (tables \ref{tab:9jcalcs} and \ref{tab:9jcalcsctd}\,), after normalization, are squares of Racah $9j$ symbols (equations (\ref{eq:9jImplicit}), (\ref{eq:ProjProd}) and Lemma \ref{eq:PtwiseNeq2Full}), which are functions of the three spin and internal labels of the two multiplicands, and the three labels of each summand.  \hfill $\Box$
\end{rem}}
\noindent
The results are tabulated in conventional angular momentum labelling rather than Dynkin labelling; specific numerical instances can be verified directly by using known tabulations, or an on-line calculator\footnote{Such as \texttt{http://www-stone.ch.cam.ac.uk/cgi-bin/wigner.cgi} by 
Anthony Stone.}\,. Note that further pointwise products for other low degree cases 
are in principle calculable using these spin chain methods, although the combinatorial manipulations will increase in complexity. For example, the product with $\Tr(u)\Tr(v)\equiv \Tr(u\otimes v)$\, is available
from Lemmata \ref{lem:TruProdForm} and \ref{lem:TrvProdForm}, while that for $\Tr(uv)= \sum \Tr(u{\sf z}\otimes v{\sf z}')$\, would necessitate rearranging the switch terms into combinations of ${\mathbb J}_{j\pm 1,1}$ and ${\mathbb J}_{j\pm 1,0}$\,, and extracting 
the corresponding projectors, finally yielding (c.f. table \ref{tab:Neq2exx}) the
expansions of \smash{$\chibig^1_{(\fh\fh)}\cdot \chibig{}_{j,\fh}^{j\pm 1/2}$} and \smash{$\chibig^0_{(\fh\fh)}\cdot \chibig{}_{j,\fh}^{j\pm 1/2}$}\,.\hfill $\Box$

\section{An application: the lattice gauge Hamiltonian.}
\label{Hamilton}


In this section we shall demonstrate how the knowledge of the multiplicative structure of 
$\mathcal R(G)$ is fundamental in Hamiltonian lattice gauge theory (with gauge group 
$
G = \SU(2)\, .
$ for this case)\,.
The material in the section is based on \cite{Fuchs2018costratificationMR,jarvis2021quasicharactersMR}\,.
Here, we limit our attention to the study of the spectral problem of the Hamiltonian. In \cite{Fuchs2018costratificationMR}, the reader may also find a detailed discussion of the role of the classical gauge orbit type  structure \cite{fuerstenberg2017defining} at the quantum level, and the associated
implications for quasicharacters.

Let $\Lambda$ be a finite spatial lattice with lattice spacing $\delta$\,, and let $\Lambda^0$, $\Lambda^1$ and $\Lambda^2$ denote, respectively, the sets of lattice sites, lattice links and lattice plaquettes. For the links and plaquettes, let there be chosen an arbitrary orientation. In lattice gauge theory, gauge potentials (the variables) are approximated by their parallel transporters along links and gauge transformations (the symmetries) are approximated by their values at the lattice sites. Thus, the classical configuration space is the space $G^{\Lambda^1}$ of mappings $\Lambda^1 \to G$, the classical symmetry group is the group $G^{\Lambda^0}$ of mappings $\Lambda^0 \to G$ with pointwise multiplication and the action of $g \in G^{\Lambda^0}$ on $\ul a \in G^{\Lambda^1}$ is given by  
\beq
\label{G-Wir-voll}
(g \cdot \ul a)(\ell) := g(x) \ul a(\ell) g(y)^{-1}\,,
\eeq
where $\ell = (x,y) \in \Lambda^1$ where $x$, $y$ denote the starting point and the endpoint of $\ell$, respectively. The classical phase space is given by the associated Hamiltonian $G$-manifold, and the reduced classical phase space is obtained from that by symplectic reduction (details are given in \cite{Fuchs2018costratificationMR,jarvis2021quasicharactersMR}). 
Dynamics is ruled by the classical counterpart of the Kogut-Susskind lattice Hamiltonian. If we identify $\ctg G$ with $G \times \mf g$, and, thus, $\ctg G^{\Lambda^1}$ with $G^{\Lambda^1} \times \mf g^{\Lambda^1}$, by means of left-invariant vector fields, the classical Hamiltonian is given by 
$$
H(a,E)
 = 
\frac{g^2}{2 \delta} \sum_{\ell \in \Lambda^1} \|E(\ell)\|^2
 -
\frac{1}{g^2 \delta} \sum_{p \in \Lambda^2} \left(\Tr a(p) + \ol{\Tr a(p)}\right)
 \,.
$$
Here, $g$ is the coupling constant, and for plaquette $p= (\ell_1,\ell_2,\ell_3,\ell_4)$, $a(p)= a(\ell_1)a(\ell_2)a(\ell_3)a(\ell_4)$, where the links $\ell_1,\ell_2,\ell_3,\ell_4$, in this order, form the boundary of $p$ and are endowed with the boundary orientation. 
{Finally, $\| \cdot \| $ denotes the norm given by the scalar product (Killing form) on $\mf g = \su(2)$.}

We now turn to the quantum Kogut-Susskind Hamiltonian $\mr H$, obtained via canonical quantization in the tree gauge (see below). The pure gauge part, treated here, reads:
\beq
\label{Hamiltonian}
\mr H = \frac{g^2}{2\delta} \, \mf C - \frac{1}{g^2 \delta} \, \mf W
\,.
\eeq
Here,
$$
\mf C := \sum_{\ell \in \Lambda^1} E_{ij}(\ell) E_{ji}(\ell)
$$
 is the Casimir operator (negative of the group Laplacian) of $\SU(2)^N$ and 
$$
\mf W := \sum_{p \in \Lambda^2} ( W (p) + W(p)^*)
\,,
$$
where the so-called Wilson loop operator $W(p)$ is the quantum counterpart of $\Tr a(p)$ (the corresponding multiplication operator on Hilbert space $\mc H$)\,. For details, see \cite{qcd3}, \cite{GR2}, \cite{KS}. Recall that the representative functions of spin $j$ on $\SU(2)$ are eigenfunctions of the Casimir operator of $\SU(2)$ corresponding to the eigenvalue\footnote{See \cite{Helgason}, \cite{Fegan}}
\beq
\label{eq:Casimir}
\epsilon_j = 4 j(j+1)
\,.
\eeq
It follows that the invariant representative functions $\BF{\ul j}{j}{\ul l,\ul l'}$ are eigenfunctions of ${\mathfrak C}$ corresponding to the eigenvalues
\beq
\label{Eigenvalue-Casimir}
\epsilon_{\ul j} = \epsilon_{j^1} + \cdots + \epsilon_{j^N}
\,. 
\eeq
Let us analyze $\mf W$. For that purpose, for our regular cubic lattice, we define a standard gauge fixing tree as follows. By a line, we mean a maximal straight line consisting of lattice links. First, choose a lattice site $x_0$ and a line $L_1$ through $x_0$. Next, choose a second line $L_2$ through $x_0$ perpendicular to $L_1$ and add all lines parallel to $L_2$ in the plane spanned by $L_1$ and $L_2$. Finally, add all lines perpendicular to that plane. Let  $B$ be such a standard tree. Since 
$a(\ell) = \mathbbm{1}$ for every $\ell \in B^1$, we can decompose $\mathfrak{W}$ into the sum of three pieces.\footnote{Note that, for the standard tree, no plaquettes 
having 3 off-tree links occur.} It is easy to check that there exists an orientation and a labelling of the off-tree links such that for every plaquette with four off-tree links (all of these plaquettes are parallel to the plane spanned by the lines $L_1$ and $L_2$), the boundary links are labelled and oriented consistently. This means that for one of the two possible orientations of the plaquette, they carry the induced boundary orientation, and that their numbers increase in that direction. Then, 
\begin{align*}
\mathfrak{W}
 =~ & 
\sum_{\{p:\,p \cap B =\emptyset\}} 
\Tr(a_{r_p} a_{s_p} a_{t_p} a_{u_p})
 +
\overline{\Tr(a_{r_p} a_{s_p} a_{t_p} a_{u_p})}
\\
 &
+ \sum_{\{p:\,|p \cap B|=2\}}\Tr(a_{r_p} a_{s_p}) + \overline{\Tr(a_{r_p} a_{s_p})}
\\
 &
+ \sum_{\{p:\,|p \cap B|=3\}}\Tr(a_{r_p})+\overline{\Tr(a_{r_p})}.
\end{align*}
To find the matrix elements of $\mr H$ with respect to the basis functions $\{\BF{\ul j}{j}{\ul l,\ul l'}$, we have to find the corresponding expansion of $\mf W$. The sequences $\ul j$ occuring here will have at most four nonzero entries, so we can write 
$\ul j = (j_1r_1,\dots,j_kr_k)$ if $\ul j$ has entries 0 except for $j_i$ at the places $r_i$, $i=1,\dots,k$. The function $T_r(\ul a) = \Tr(a_r)$ coincides with the basis function $\BF{\ul j}{j}{\ul l,\ul l'}$ with $\ul j = (\frac12 r)$. If we omit the irrelevant indices 
$j , \ul l , \ul l'$, we thus have 
$$
T_r = \BFi{(\frac12 r)}
\,.
$$
The function $T_{rs}(\ul a) := \Tr(a_r a_s)$ is a linear combination of the basis functions $\BF{\ul j}{j}{\ul l \ul l'}$ with $\ul j = (\frac12r,\frac12s)$. Again, we may omit the irrelevant labels $\ul l$, $\ul l'$. 
Now, using Proposition \rref{S-chi-D}, we can write
\beq\label{G-chi-MEF-2}
\BF{(j_rr,j_ss)}{j}{}
 =
\sqrt{\frac{d_{j_r} d_{j_s}}{d_j}} 
 ~
\sum\limits_{m=-j}^j
 ~
\sum\limits_{n_r + n_s = m \atop n_r' + n_s' = m}
 ~
C^{j_r j_s j}_{n_r n_s m} C^{j_r j_s j}_{n_r' n_s' m}
 \,
D^{j_r}_{n_r n_r'}(a_r) 
 \,
D^{j_s}_{n_s n_s'}(a_s) \, .
\eeq
Using this, together with the orthogonality relation 
\beq\label{G-OR}
\braket{D^j_{m_1 m_2}}{D^{j'}_{m_1' m_2'}}
 = 
\frac{1}{d_j} \, \delta_{jj'} \delta_{m_1 m_1'} \delta_{m_2 m_2'}
\eeq
and the normalization condition $\sum_{m_1,m_2=\pm\frac12} (C^{\frac12 \frac12 j}_{m_1 m_2 m})^2 = 1$, we obtain
$$
\braket{\BF{(\frac12r,\frac12s)}{j}{}}{T_{rs}}
 =
(-1)^{1-j} \, \frac{\sqrt{d_j}}{2}\,.
$$
Thus,
\beq
\label{eq:TrsExp}
T_{rs}
 = 
\frac{\sqrt 3}{2} \, \BF{(\frac12 r,\frac12 s)}{1}{}
 -
\frac 1 2 \, \BF{(\frac12 r,\frac12 s)}{0}{}
\,.
\eeq
Finally, the function $T_{rstu}(\ul a) := \Tr(a_r a_s a_t a_u)$ is a linear combination of the basis functions $\BF{\ul j}{j}{\ul l \ul l'}$ with $\ul j = (\frac12r,\frac12s,\frac12t,\frac12u)$. Here, the sequences $\ul l \in R(\ul j,j)$ have entries $l^1 = \cdots = l^{r-1} = 0$, $l^r = \cdots = l^{s-1} = \frac12$, $l^s = \cdots = l^{t-1} = l$, $l^t = \cdots = l^{u-1} = k$ and $l^u = \cdots = l^N = j$, where $l =0,1$ and $k \in \langle l , \frac12 \rangle$ so that $j \in \langle k , \frac12 \rangle$. This means that  they are labelled by two intermediate spins $l,k$, so that in our notation we may replace the labels $\ul l$ and $\ul l'$ by $(l,k)$ and $(l',k')$, respectively. Expressing these basis functions in terms of matrix elements according to Proposition \rref{S-chi-D}\,, and using once again the orthogonality relation \eqref{G-OR}, we obtain
 \ala{
\braket{\BF{(\frac12r,\dots,\frac12u)}{j}{(l,k),(l',k')}}{T_{rstu}}
 & =
\frac{1}{4 \sqrt{d_j}}
 \,
\sum_{m=-j}^j ~ \sum_{m_r+\cdots+m_u=m} ~ 
\\
 & \hspace{1cm}
C^{\frac12 \frac12 l}_{m_r,m_s,m_r+m_s}
C^{l \frac12 k}_{m_r+m_s,m_t,m_r+m_s+m_t}
C^{k \frac12 j}_{m_r+m_s+m_t,m_u,m}
\\
 & \hspace{1cm}
C^{\frac12 \frac12 l'}_{m_u,m_r,m_r+m_u}
C^{l' \frac12 k'}_{m_r+m_u,m_s,m_r+m_s+m_u}
C^{k' \frac12 j}_{m_r+m_s+m_u,m_t,m}
\,.
 }
A detailed calculation (see \cite{jarvis2021quasicharactersMR}) leads to 
\begin{align}
\label{eq:TrstuExp}
 T_{rstu}
 =&\, \textstyle \frac 1 8 \BF{(\frac12r,\cdots,\frac12u)}{0}{(0\frac12)(0\frac12)}
 -\frac{\sqrt{3}}{8} \BF{(\frac12r,\dots,\frac12u)}{0}{(0\frac12)(1\frac12)}
-\frac{\sqrt{3}}{8} \BF{(\frac12r,\dots,\frac12u)}{0}{(1\frac12)(0\frac12)}
-\frac{1}{8} \BF{(\frac12r,\dots,\frac12u)}{0}{(1\frac12)(1\frac12)}\nonumber \\ 
&\,\textstyle-\frac{\sqrt{3}}{8} \BF{(\frac12r,\dots,\frac12u)}{1}{(0\frac12)(0\frac12)}
+\frac 3 8 \BF{(\frac12r,\dots,\frac12u)}{1}{(0\frac12)(1\frac12)}
+ 0 \, \BF{(\frac12r,\dots,\frac12u)}{1}{(0\frac12)(1\frac32)}\nonumber \\ 
&\, \textstyle-\frac 1 8 \BF{(\frac12r,\dots,\frac12u)}{1}{(1\frac12)(0\frac12)}
-\frac{1}{8\sqrt{3}} \BF{(\frac12r,\dots,\frac12u)}{1}{(1\frac12)(1\frac12)}
+\frac1{\sqrt{6}} \BF{(\frac12r,\dots,\frac12u)}{1}{(1\frac12)(1\frac32)}\nonumber \\ 
&\,\textstyle-\frac{1}{\sqrt{8}} \BF{(\frac12r,\dots,\frac12u)}{1}{(1\frac32)(0\frac12)}
-\frac{1}{2\sqrt{6}} \BF{(\frac12r,\dots,\frac12u)}{1}{(1\frac32)(1\frac12)}
-\frac{1}{4\sqrt{3}} \BF{(\frac12r,\dots,\frac12u)}{1}{(1\frac32)(1\frac32)}\nonumber \\ 
&\,\textstyle+\frac{\sqrt{5}}{4} \BF{(\frac12r,\dots,\frac12u)}{2}{(1\frac32)(1\frac32)}
\,.
\end{align}
\begin{rem}
The relation between the quasicharacters contributing in equations (\ref{eq:TrsExp}) and 
(\ref{eq:TrstuExp}), and the various trace types occurring, is seen by reference to the evaluations for $N=2$ and $N=4$ carried out in Section \ref{sec:SpinChainsProjectors} above.
For example, (\ref{eq:TrsExp}) follows directly from the $N=2$ table \ref{tab:Neq2exx} after normalization; 
see Example \ref{eq:Quasi111123234} for explicit evaluation of the $N=4$ quasicharacter
in (\ref{eq:TrstuExp}) with maximal spin and internal labels, namely 
${}^j(l,k)(l',k') ={}^2(1,\tfrac 32) (1,\tfrac 32)$\,.

An alternative strategy for the $T_{rstu}$ expansion would be to exploit Lemma \ref{lem:PtWiseT1T1Split} (for quasicharacter products with distinct group elements), to build iteratively up to $N=4$\,. While such products rely only on the angular momentum Clebsch-Gordan series (with unit $3j$ coefficients), the resultant expansion would then contain quasicharacters based on different tree labelling, necessitating further rearrangement (Lemma \ref{lem:TreeChange}) to arrive at an expansion such as (\ref{eq:TrstuExp}) over an orthonormal basis.  \mbox{}\hfill $\Box$
\end{rem}

Now, consider the eigenvalue problem for $\mr H$. For that purpose, we simplify the notation by collecting the 
labels $\ul j$, $j$, $\ul l$ and $\ul l'$ of the basis functions in a multi-index
$$
I :=\left(\ul j ; j ; \ul l ; \ul l'\right).
$$
Let $\mc I$ denote the totality of all these multi-indices. According to Proposition \rref{Gen-MultLaw}, the structure constants of multiplication, defined by 
\beq
\label{StructConst}
\BFCi{I_1} \cdot \BFCi{I_2}
 = 
\sum_{I \in \mc I} ~ C^I_{I_1 I_2} \, \BFCi I
\,,
\eeq
are given by 
\beq\label{Gen-structC}
C^{I}_{I_1 I_2}
 ~=~
\sqrt{\frac{d_{\ul j_1} d_{\ul j_2} d_j}{d_{j_1} d_{j_2} d_{\ul j}}}
 ~
U_{\ul j_1,\ul j_2}(\ul j,\ul l;\ul l_1,\ul l_2)
 ~
U_{\ul j_1,\ul j_2}(\ul j,\ul l';\ul l_1',\ul l_2')
\,,
\eeq
where $I_i = \left(\ul j_i;j_i;\ul l_i;\ul l_i'\right)$ and $I = \left(\ul j;j;\ul l;\ul l'\right)$.  Expanding
\beq
\label{Dec-mcW}
\psi = \sum_J \psi^J \BFi J
 \,,\qquad
{\mathfrak W} = \sum_I W^I (\BFi I + \ol{\BFi I})
 \,,
\eeq
and using \eqref{StructConst}, as well as the fact that $\braket{\BFi K}{\ol{\BFi I}\BFi J} = \braket{\BFi I \BFi K}{\BFi J} = C_{IK}^J$ implies
$$
\overline{\BFi I} \BFi J = C_{IK}^J \BFi K
\,,
$$
Finally, the eigenvalue problem reads 
\beq
\label{EP-H}
\sum_{J \in \mc I} 
 \left\{ \left( 2{g^2}{\delta} \epsilon_J - {\cal E} \right) \delta_J^K
 - 
\frac{1}{g^2 \delta} \sum_{I \in \mc I} W^I
\left(C_{IJ}^K + C_{IK}^J\right)\right\} \psi^J
 = 
0
 \,,
\eeq
for all $K \in \mc I$. Here, $\epsilon_J$ stands for  the eigenvalue of the Casimir operator $\mf C$ corresponding to the eigenfunction 
$\BFi J$, given by \eqref{Eigenvalue-Casimir}. Thus, we have obtained a homogeneous system of linear equations for the eigenfunction coefficients $\psi^J$. The eigenvalues $\cal E$ are determined by the requirement that the determinant of this system must vanish. Note that the sum over $I$ in \eqref{EP-H} is finite, because there are only finitely many nonvanishing $W^I$. Moreover, by Remark 4.12 in \cite{Fuchs2018costratificationMR}, also the sum over $J$ is finite for every fixed $K$. Thus, we have reduced the eigenvalue problem for the Hamiltonian to a problem in linear algebra. Combining this with well-known asymptotic properties of $3nj$ symbols (see \cite{biedenharn1981angular} (Topic 9),\cite{Anderson}, and further references therein\,), we obtain an algebraic setting which allows for a computer algebra supported study of the spectral properties of $H$.

\section{Conclusions.}
\label{sec:Concl}
In the present study we have developed an analysis of the algebra ${\mathcal R}(G)$ of $G$-invariant representative functions over $N$ copies of a group $G$, adapted to the case of angular momentum, $G=\SU(2)$\,.
The property of invariance under simultaneous conjugation by a fixed goup element 
is a generalization of the $N=1$ case, which characterizes the standard theory of group characters. We have introduced the corresponding generalized irreducible 
quasicharacters, and for $G=\SU(2)$ we have provided concrete methods for 
computations in examples of low degree - both explicit evaluations, and product expansions(sections \ref{subsec:ExplicitNeq2n3characters} and \ref{subsec:ExplicitNeq2products}\,). The existence of orthonormal basis sets guaranteed by the group invariant measure, leads to a characterization of equivalent 
bases, arising via traces of group element representatives over totally coupled states. 
As discussed above, the expansion of pointwise products, as well as changes of coupling scheme, or tree (see Appendix), lead to a distinguished role for Racah recoupling coefficients as overlap and multiplicative structure constants. In particular, 
the latter are shown to be $3(2N-1)j$ Racah symbols, generalizing the $N=1$ case
where the $3j$ symbols simply specify the angular momentum triangle condition (sections \ref{subsec:Properties}\,,\ref{subsec:ExplicitNeq2Products9j}\,). 

We refer to table \ref{tab:MonomialCount} for an examination of the combinatorial stucture of the invariant ring ${\mathcal R}(G)$ in the $\chibig^T(\ulu)$ basis in more detail. The column sums (the Catalan sequence) reflect the overall multiplicity of total angular momenta obtained in the tensor product of $N$ copies of the fundamental spin-$\textstyle{\frac 12}$ representation (and coincide with the squared sum $\sum_j m_{\ulj}(j)^2$ of multiplicities in the reduction of each
module $V_{\ulj}$\,, by bracketing the tensor products of spin-$\textstyle{\frac 12}$ modules (as in equation (\ref{G-D-Hjl}) ) and assembling the resulting $V_{\ulj,j}$ modules\,). The list entries refer to monomials in trace strings which are in bijection with the total count. For example, for $N=2$ we have $\{ Tr(uv),
Tr(u)Tr(v)\} $\,, as well as $\{ Tr(u)Tr(v)Tr(w), Tr(uv) Tr(w),Tr(vw) Tr(u),Tr(wu) Tr(v),$ $Tr(u[v,w])\}$\,
for $N=3$\,. For degree higher than 5, however, the entries marked ${}^*$ represent reduced counts modulo additional relations which hold (including the restriction to
partitions with parts of length $\le 3$), pertaining to the generating sets of various matrix invariant rings (see \cite{Procesi73}  and for example \cite{Formanek1984}, \cite{Drensky2007})\,.
In general, however, it is evident that the entries in tables \ref{tab:Neq2exx} and 
\ref{tab:Neq3exx}
represent Kostka-type transformation coefficients between the orthonormal basis of 
irreducible quasicharacters, and selected monomial basis representatives. 
 
In the present context, the well known association between the recoupling calculus, and tensor categories of group representations, thus becomes mirrored in the role of  the recoupling calculus in supporting the invariant \emph{ring} of generalized group characters. Even for the $N=1$ case and $G=\SU(2)$ (angular momentum),  this can be seen in the associativity
$\langle \langle j_1, j_2\rangle, j_3\rangle =
\langle  j_1,\langle j_2, j_3\rangle\rangle $ of the set-theoretical operator specifying
the angular momentum combination rules (equivalently, associativity of the $3j$ structure constants governing pointwise products of irreducible characters).
On the other hand, from the pointwise product Lemma, while not manifest, the presence of recoupling \emph{coefficients} as the $N\ge 2$ structure constants ensures that associativity (and of course commutativity) is still consistent. This is clearly a concomitant of the tensor-categorical origins of the recoupling calculus itself \cite{joyce2002racah}. 

It should further be noted that the general representation-theoretical setting for 
quasicharacters, namely, an $N$-fold tensor product of $G$-modules, is precisely the arena in which the so-called quadratic Racah algebra $R(n)$ is defined
(see for example \cite{deBie2017higher}\,). In this case the 
algebraic generators are the partial Casimir operators \smash{${C}_{ij}:= (\bfJ_i\!+\! \bfJ_j)^2$} rather than the projection operators \smash{$\bP_{j_ij_j}^j$} which arise in the spectral decomposition \smash{${C}_{ij}= {\sum}_j j(j\!+\!1)\bP_{j_ij_j}^j$}. The rule (\ref{eq:ProjProd}) for expanding projection operator products, with the expansion coefficients being $9j$ Racah symbols, is thus an aspect of fine-grained
structure wherein, in this instance, underlying the general $R(n)$ quadratic bracket algebra, certain operator products themselves can be evaluated explicitly in closed form. 

There appears moreover to be a functorial association between the pointwise algebraic product, and the combinatorial tree operations of join and multiply. For example, while
products of more than two quasicharacters can by definition be evaluated pairwise,
it is natural to conjecture that the resultant expansions can be recombined into
a final summation reflecting the tree structure. In this way, the binary multiplication, schematically,
\[
\chibig^T \!\cdot\! \chibig^T
= \sum R\big((T\!\cdot\! T)| T^{\scalebox{.8}{\bVee}}\big)^2
\chibig^T\,, 
\]
would have a ternary generalization
\[
(\chibig^T \!\cdot\! \chibig^T)\!\cdot\! \chibig^T
= \sum R\big((T\!\cdot\! T)\!\cdot\!T| T^{\scalebox{.8}{\bWee}}\big)^2
\chibig^T\,, 
\]
(with a similar expression for the opposite bracketing given that the associator must vanish), and also possibly for higher order products.

Additional structure in the case of the angular momentum recoupling calculus, includes
the classification of fundamental recouplings 
according to equivalence classes of certain maximally connected trivalent graphs \cite{Yutsis,louck2008unitary} associated with the coupling trees involved
\footnote{For one-or two-line connected graphs, the associated recouplings are
factorizable. Thus, whereas there is only one type of $6j$ or $9j$ Racah symbol, there are respectively two, and five, fundamental types for $12j$ and $15j$\,.}\,.  It is a natural question to investigate how these distinctions influence specific quasicharacters, in their behaviour as multiplicands. 
It can also be expected that the functorial correspondences between algebraic operations on characters, and combinatorial moves on trees, emerging from our preliminary analysis, can be given more formal foundations.
All of these considerations have relevance to the motivation of the applications in Hamiltonian lattice gauge theory and geometric quantization, and will form topics for ongoing study. In  the case of physical gauge group $G=SU(3)$, the recoupling theory must take account of the non-multiplicity free nature of the tensor products (see \cite{jarvis2021quasicharactersMR}\,). We hope to return to these questions in future work. 
\mbox{}
\\[1cm]

\noindent
\textbf{Acknowledgements}\\
The authors thank Ronald King for discussions and valuable suggestions on the work.


\begin{appendix}
\renewcommand{\theequation}{A--\arabic{equation}}
\section{$\SU(2)$ irreducible quasicharacters and angular momentum recoupling calculus.}
\label{sec:IrrGchandRecoupling}
In the foregoing the ring ${\mathcal R}(\SU(2))$ of $\SU(2)$-invariant representative functions associated to the irreducible quasicharacters has been analyzed in the context of related combinatorial results in invariant theory, especially as regards counts and gradings. It is natural also to review the algebraic structure in more detail. This is the focus of the present section, bringing together various results that have been povided in 
\cite{Fuchs2018costratificationMR,jarvis2021quasicharactersMR}\,.
While the $G=\SU(2)$ case is our primary concern here, some of the discussion nonetheless applies more generally, and this will be indicated at the appropriate places.

In this appendix we
replace the label set $\underline{j}, \underline{\ell}$ by the data 
 of a \emph{binary coupling tree}.
The choice of (planted), semi-labelled binary coupling tree and the composite data $T, \underline{j}, \underline{k}, j$\, (the notation will be explained presently)\,, will turn out to be key to unravelling the algebraic structure of the group invariant representative functions ${\mathcal R}(\SU(2))$ in terms of concrete manipulations with the associated irreducible quasicharacters $\chibig$, via the dependence on $T$ and transformations between different coupling types.
We therefore preface the discussion here with some notational conventions regarding tree and list manipulation. In the following subsections, the behaviour of the $\chibig$ characters under change of $T$, and also their structure as a commutative algebra (pointwise multiplication), will be described.

\subsection{Notation.}
\label{subsec:Notation}
An irreducible representation $V_{\ulj}$ of the direct product group $\times^N\!\SU(2)$ 
admits a complete reduction under the diagonal subgroup $\SU(2)^d$ into a direct sum 
$
V_{\ulj} \cong \left.\sum\right._j \oplus  m_{\ulj}(j)V_j
$
wherein the isocopy decomposition  
$V_{\ul j,j} = V_j \oplus V_j \oplus \cdots \oplus V_j $ (for multiplicity $m_{\ulj}(j) >1$) is defined only up to isomorphism,
for example dependent on a choice of intermediate submodules of 
coupled tensor products of elements of $V_{\ul j}$ and an associated descending chain of subgroups  
(see equations (\ref{G-D-Hjl}) and (\ref{eq:RjDef})\,). Explicit construction of such couplings requires adoption of explicit bases, and use of the appropriate \emph{coupling coefficients} as discussed in the main text. 
 The decomposition into irreducible subspaces can be cast into more combinatorial terms as follows.
\begin{quotation}
\mbox{}

\noindent
\textbf{Coupled states}\\
Given a planted binary tree $T$ with $N$ numbered leaf nodes. Let $(i,i')$ be the system of pairings belonging to the associated perfect matching
\cite{erdHos1989applications},\cite{diaconis1998matchings},\cite{francis2022brauer}, ordered from 1 to $N-1$ via the bijective correspondence which defines the tree, with $i,i' \in [2N-2]$\,, corresponding edge decorations $(s_i, s_i')$ (including the given angular momentum leaf edge labels), such that each child node $i''$ has edge decoration $s_i''\in \langle s_i, s_i'\rangle$\,. Node labelled $2N-1$, the root, has parents $(s_r, s_r')$ such that for the root edge\,, $j\in \langle s_r, s_r'\rangle$\, (but the pair $(r, r')$ does not belong to the matching\,). 
Then the angular momentum coupled states are (compare equation (\ref{eq:uljullmCoupled})\,)
\begin{align}
 \label{eq:GenCGcoeffs}
| T(\ulj \ul k) jm \rangle =&\, \sum_{\ul m} |\ul j, \ul m\rangle
\langle \ul j, \ul m |T(\ulj \ul k) jm \rangle\,, \quad \mbox{where}  \nonumber \\
\langle \ul j, \ul m |T(\ulj \ul k) jm \rangle = &\,\Big(\prod_{(s,s'),m,m'}\,C^{s,s',s''}_{m,m',m''} \Big)
C^{s_r,s_r',j}_{m_r,m_r',m}\,.
\end{align} 
with relabelling of the edge decoration list as $\ul j, \ul k, j$ according to a
fixed presentation of $T$\,\footnote{The list $k_1, k_2,\cdots$ on a labelled tree diagram is mapped uniquely to edge decorations $s_{N+1}, s_{N+2}, \cdots$ of the ordered node list $N+1,N+2,\cdots$ via the bijection algorithm.
In this presentation there is the analogue of the subgroup chain 
(equation (\ref{eq:SubgroupChaindDef})) and associated sets
$R_{\ul j}(T)$\,, $R_{\ul j}(T,j)$ (equation (\ref{eq:RjDef})) adapted to the binary tree and ordered pairings accorded by the perfect matching, together with projectors 
$\bP(T)^j_{\ul j, \ul k}$ (c.f. equation (\ref{Dec-ProjOp})\,). }\,. \mbox{}\bx
\end{quotation}
%
\noindent
We now introduce the basic quantities which underly the unitary equivalence of 
the admissible tensor product decompositions.
\begin{quotation}
\mbox{}

\noindent
\textbf{Recoupling coefficients \cite{biedenharn1981racah}}\\
For two different labelled trees (coupling schemes) $T, T'$ the overlap of the coupled states, 
$\langle T'(\ulj \ulK) J M | T(\ulj \ulK) jm \rangle $ is zero if $J\ne j$ and $M\ne m$\,, and moreover by covariance independent of $m$. The quantities
\begin{equation}
\label{eq:RacahCoeffDef}
\big( R_{T'T}\big)^{\ulj}_{\ulK j, \ulk j} = \langle T'(\ulj \ulK) j m | T(\ulj \ulK) jm \rangle
\end{equation}
are the angular momentum \emph{recoupling coefficients} (Racah coefficients). \hfill $\Box$
\end{quotation}
\noindent
In this general case the recoupling coefficients are of type $3N-3$ as they are functions of the coupled state vectors with $N+1$ fixed labels $\ulj, j$\,, and the additional $2(N-2)$ intermediate labels $\ulK,\ulk$. They form a unitary array of size $m_{\ul j}(j) \times m_{\ul j}(j)$ which effect the isomorphism amongst isocopy spaces for given $j$, for varying couplings.
We define two operations on trees \cite{LodayArith2002} and corresponding coupling trees :
\begin{quotation}
\noindent
\textbf{Tree join, $T_{\bcdot} =T\bcdot T'$:}\\
The join $T_{\bcdot} =T \bcdot T'$ is the tree with $|L_{T_{\bcdot}}|= 2N$ leaves formed by glueing the root edges
of $T$ and $T'$ to make a new internal node, whose out edge terminates in the new root.
The leaf labelling is thus the join 
\[
\ulj\cdot \ulj' = (j_1, j_2,\cdots, j_N, j'_1, j'_2\cdots, j'_N)
\]
of the leaf sets, and the internal edge labels also incorporate the total angular momenta $j$ and $j'$ of $T$ and $T'$\,, which combine to a total angular momentum $J$. In the following we use adaptations of this notation for other label sets; for example
$\ulm_1\cdot \ulm_2\,, \ulk_{1\cdot 2}$\,. \\[.3cm]
\textbf{Tree composition, $T^{T'} =T \ast  T' $:}\\  
The product $T^{T'} =T \ast  T' $ is the tree with $|L_{T_\ast}| = |L_T||L_{T'}|$ leaves
formed by replacing each leaf edge of $T$ by the root edge of a copy of $T'$ above it.\\[.3cm]
and in particular we have\\[.3cm]
\textbf{Leaf duplication, $T^{\scalebox{.8}{$\bVee$}}=T \ast  \bVee $:}\\ 
The coupling tree formed by multiplication by a single cherry $\bVee$, thus has $2N$ leaves, which can be assigned labels as the thread
\[
\ulj \ast \ulj' = (j_1,j'_1, j_2,j'_2, \cdots, j_N, j'_N)
\]
of two different leaf sets $\ulj, \ulj'$. In $T^{\scalebox{.7}{$\bVee$}}$\,, the original leaf labels $\ulj$ and $\ulj'$ of $T$, now become coupled angular momenta $j_{11'}, j_{22'}, \cdots$\,, corresponding to couplings of $j_1$ with $j_1'$\,, $j_2$ with $j_2'$\,,
$\cdots$\,, and the final total angular momentum $J$ replaces $j$ on the edge correspondng to the root of $T$\,.  In the following we use adaptations of this notation for other label sets; for example
$\ulk_1\ast \ulk_2\,, \sum_{\ulm^{1\ast2}}\,, \cdots$\,. 
\end{quotation}

\subsection{Properties: change of coupling tree
$\langle \chi^{\widetilde{T}}_{j} |\chi^T_j\rangle$, and pointwise multiplication
$\langle \chi^{T}_{j''}|\chi^T_j\!\cdot \! \chi^{T}_{j'}\rangle$\,.}
\label{subsec:Properties}

Recall that the irreducible quasicharacters are specified by sums over total angular momentum matrix elements between coupled states constructed using a given coupling tree, but possibly different internal labels:
\[
{\chibig}^T_{(\ulj)\ulk \ulk'; j}(\ulu)
= \sum_m \langle T(\ulj) \ulk; jm| D_{\ulj}(\ulu)
| T(\ulj) \ulk'; jm \rangle\,.
\]
Inserting complete sets of coupled states for a second coupling tree $\widetilde{T}$\,, and rearranging, leads to the 
\begin{lem}
[Transformation rule under change of coupling tree]
\label{lem:TreeChange}
\begin{align*}
\chibig^T_{(\ulj)\ulk \ulk';j}(\ulu)
= &\, \sum_{J,\ulK,\ulK'}  
R\big(T|\widetilde{T}\big)^{\ulj}_{\ulk j \ulK j} \,
   \chibig^{\widetilde{T}}_{(\ulj)\ulK \ulK';j}(\ulu)\, R\big(\widetilde{T}|{T}\big)^{\ulj}_{\ulK'j \ulk'j} \, .
\end{align*}
\mbox{}\hfill $\Box$
\end{lem}

Evaluation of pointwise products of general quasicharacters proceeds via
a schematic coupling and uncoupling manipulation. On the one hand, the product of 
traces can be viewed as a trace over a tensor product, for which totally
coupled states are associated with the join $T\cdot T$ of the individual coupling trees.
On the other hand, the coupling tree $T$ is a subtree of the thread $T \ast \bVee$,
where the $2N$ angular momentum labels $\ulj$ and $\ulj'$ must also be reordered
as $\ulj \ast \ulj'$ leaf labels, with the coupled $N$-component $\ulJ$ in turn
providing the leaf labels of the subtree $T$:
\begin{lem}
[Pointwise product of quasicharacters]
\label{lem:PtWiseT1T2}
\begin{align*}
\chibig^{T,j_1}_{\ulj,\ulk_1\ulk'_1}(\ulu) & \cdot \chibig^{T,j_2}_{\ulj,\ulk_2,\ulk'_2}(\ulu)  \nonumber\\
& \, =
\sum_{j, \ulJ\ulK ,\ulK' }
\big(R^{\ulj\ast \ulj}_{ T\cdot T,T\ast \scalebox{.8}{$\bVee$}}\big)_{\ulk_{1\cdot2},j_1 j_2,j; \ulJ,\ulK,j }
\big(R^{\ulj\ast \ulj}_{T\ast \scalebox{.8}{$\bVee$},T\cdot T}\big)_{ \ulJ,\ulK',j ; \ulk_{1'\cdot2'},j_1 j_2,j} \cdot \chibig^{T,j}_{\ulJ, \ulK,\ulK'}(\ulu)\,.
\end{align*}
\mbox{}\hfill $\Box$
\end{lem}

In this case the recoupling coefficients measure the overlap of two $2N$-leaf coupling trees, and so each entail $2(2N)-1$ labels; discounting for each the $N$ repeated spin labels within $\ulj \ast \ulj$, and the overall common $j$, they are thus of $(6N-3)j$- type, as appropriate for structure constants arising from expanding the pointwise product of two $\chibig$ characters into a sum, $3(2N-1) = 6N-3$. 
The structure constants reflect the commutativity of the pointwise product because of 
their symmetry (under interchange of $\ulk_1, \ulk'_1$ and $\ulk_2, \ulk'_2$ 
and $\ell_1, \ell_2$). (As shown in Lemma \ref{lem:PtWise9jFactors}
(see \cite{Fuchs2018costratificationMR,jarvis2021quasicharactersMR}), the recoupling coefficients factorize as a product of $9j$ symbols).

For pointwise multiplication $\chibig^{T_1}\bcdot \chibig^{T_2}$ of quasicharacters from two different coupling trees, the above formalism also goes through, and indeed the result can be expressed in the $\chibig^{T_3}$ basis provided by a third tree (either directly, or via the transformation rule under change of coupling tree).

An important instance of pointwise multiplication is for a
quasicharacter product 
\[
\chibig(T_1)_{(\ul j^{(1)})\ul k^{(1)}\ul k'{}^{(1)}}(\ul u^{(1)})\cdot
\chibig(T_2)_{(\ul j^{(2)})\ul k^{(2)}\ul k'{}^{(2)}}(\ul u^{(2)})
\]
of type $N_1$ with type $N_2$\,, with distinct group elements 
$(u^{(1)}_1, u^{(1)}_2, \cdots, u^{(1)}_{N_1})$ and
$(u^{(2)}_1, u^{(2)}_2, \cdots, u^{(2)}_{N_2})$\,, respectively.
This can be treated using Lemma \ref{lem:PtWiseT1T2} above,
regarding each quasicharacter to be of type $N_1+N_2$ by expanding the
spin labels of each multiplicand, namely $(\ul j^{(1)}, \ul 0)$,
and $(\ul 0, \ul j^{(2)})$\,, respectively. However, as usual, 
quasicharacter products may be written as the trace over the  
tensor product of the appropriate projectors in the total space. This tensor product is the penultimate pairing in the sequence of 
projectors based on the tree $T=T_1 \cdot T_2$\,, whose
previous couplings operate entirely separately on each subspace $H_{\ul j^{(1)}}\otimes
H_{\ul j^{(2)}}$, corresponding to the left and right subtrees (see section\footnote{The corresponding perfect matching is itself the union of matchings for $T_1$ and $T_2$\,.}
\ref{subsec:Notation} and equations (\ref{G-D-Hjl}), (\ref{eq:SubgroupChaindDef})\,.
Thus we can write, for the final pairing
\[
H_{\ul j^{(1)}}\otimes H_{\ul j^{(2)}} \cong
\sum_{J \in \langle{j_1,j_2}\rangle }\!\! \!\boplus\, H_J 
\]
and correspondingly
\[
\bP(T_1)_{(\ul j^{(1)})\ul k^{(1)}}\otimes\bP(T_2)_{(\ul j^{(2)})\ul k^{(2)}} =
\sum_{J \in \langle{j_1,j_2}\rangle }\bP^J_{j_1,j_2}
\big(\bP(T_1)_{(\ul j^{(1)})\ul k^{(1)}}\otimes\bP(T_2)_{(\ul j^{(2)})\ul k^{(2)}})\,.
\]
The manipulation on each pair $\ul k^{(1)}, \ul k'{}^{(1)}, j^{(1)}$,
$\ul k{}^{(2)}, \ul k'{}^{(2)}, j^{(2)}$ of degenerately labelled subspaces
leads to

\begin{lem}[Pointwise product with independent group elements] 
\label{lem:PtWiseT1T1Split}
\[
\chibig(T_1)^{j^{(1)}}_{(\ul j^{(1)})\ul k^{(1)}\ul k'{}^{(1)}}(\ul u^{(1)})\cdot
\chibig(T_2)^{j^{(2)}}_{(\ul j^{(2)})\ul k^{(2)}\ul k'{}^{(2)}}(\ul u^{(2)})
=
\sum_{J \in \langle j^{(1)},j^{(2)}\rangle}
\chibig(T_1\cdot T_2)^J_{(\ul j^{(1)}\ul j^{(2)}) \ul K, \ul K' }
\]
with internal edge labels $\ul K = (\ul k^{(1)},\ul k{}^{(2)}, j^{(1)},j^{(2)} )$
and $\ul K' = (\ul k{}^{(2)},\ul k'{}^{(2)}, j^{(1)} j^{(2)})$\,.
\\ \bx
\end{lem}

\subsection{Products of $N=2$ quasicharacters and $9j$ symbols.} 
\label{subsec:ExplicitNeq2Products9j}

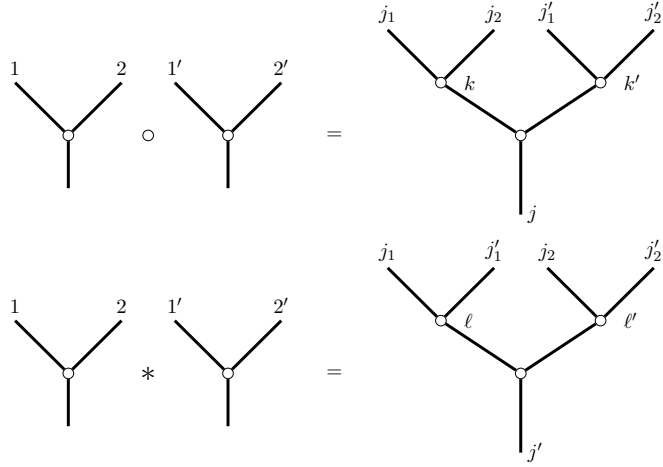
\begin{figure}[htbp]
\mbox{}\mbox{}\hskip3cm
\scalebox{.7}{\begin{tikzpicture} 
%
%
2. \draw[ultra thick] (1,3) --  (2,2) -- (3,3);
2. \draw[ultra thick] (2,1) --  (2,2);
4. \draw[fill, color=white] (2,2) circle (.1);
4. \draw (2,2) circle (.1);
2. \draw[ultra thick] (4,3) --  (5,2) -- (6,3);
2. \draw[ultra thick] (5,1) --  (5,2);
4. \draw[fill, color=white] (5,2) circle (.1);
4. \draw (5,2) circle (.1);
4. \draw[fill, color=white] (3.5,2) circle (.1);
4. \draw (3.5,2) circle (.1);
2. \draw[ultra thick] (8,4) --  (9,3) -- (10,4);
2. \draw[ultra thick] (11,4) --  (12,3) -- (13,4);
2. \draw[ultra thick] (9,3) --  (10.5,2) -- (12,3);
2. \draw[ultra thick] (10.5,.5) --  (10.5,2);
4. \draw[fill, color=white] (9,3) circle (.1);
4. \draw (9,3) circle (.1);
4. \draw[fill, color=white] (12,3) circle (.1);
4. \draw (12,3) circle (.1);
4. \draw[fill, color=white] (10.5,2) circle (.1);
4. \draw (10.5,2) circle (.1);
1. \node at (7,2) {$=$};
\node[above] at (8,4) {$j_1$};
\node[above] at (10,4) {$j_2$};
\node[above] at (11,4) {$j_1'$};
\node[above] at (13,4) {$j_2'$};
\node[right] at (9.3,3) {$k$};
\node[right] at (12.3,3) {$k'$};
\node[right] at (10.5,.5) {$j$};
\node[above] at (1,3) {$1$};
\node[above] at (3,3) {$2$};
\node[above] at (4,3) {$1'$};
\node[above] at (6,3) {$2'$};
\end{tikzpicture}}\mbox{}\\\mbox{}\hskip3cm
\scalebox{.7}{\begin{tikzpicture} 
%
%
2. \draw[ultra thick] (1,3) --  (2,2) -- (3,3);
2. \draw[ultra thick] (2,1) --  (2,2);
4. \draw[fill, color=white] (2,2) circle (.1);
4. \draw (2,2) circle (.1);
2. \draw[ultra thick] (4,3) --  (5,2) -- (6,3);
2. \draw[ultra thick] (5,1) --  (5,2);
4. \draw[fill, color=white] (5,2) circle (.1);
4. \draw (5,2) circle (.1);
1. \node at (3.5,2) {\scalebox{1.5}{$\ast$}};
2. \draw[ultra thick] (8,4) --  (9,3) -- (10,4);
2. \draw[ultra thick] (11,4) --  (12,3) -- (13,4);
2. \draw[ultra thick] (9,3) --  (10.5,2) -- (12,3);
2. \draw[ultra thick] (10.5,.5) --  (10.5,2);
4. \draw[fill, color=white] (9,3) circle (.1);
4. \draw (9,3) circle (.1);
4. \draw[fill, color=white] (12,3) circle (.1);
4. \draw (12,3) circle (.1);
4. \draw[fill, color=white] (10.5,2) circle (.1);
4. \draw (10.5,2) circle (.1);
1. \node at (7,2) {$=$};
\node[above] at (8,4) {$j_1$};
\node[above] at (10,4) {$j_1'$};
\node[above] at (11,4) {$j_2$};
\node[above] at (13,4) {$j_2'$};
\node[right] at (9.3,3) {$\ell$};
\node[right] at (12.3,3) {$\ell'$};
\node[right] at (10.5,.5) {$j'$};
\node[above] at (1,3) {$1$};
\node[above] at (3,3) {$2$};
\node[above] at (4,3) {$1'$};
\node[above] at (6,3) {$2'$};
\end{tikzpicture}}
\caption{Illustrating the join and composition of two two-leaf trees and their
associated label sets $\ulj\cdot \ulj'$ and $\ulj\ast\ulj'$\,.}
\label{fig:4leaf9j}
\end{figure}

In the $N=2$ case both the tree join, and multiplication (leaf duplication), of two 2-leaf trees (planted cherries), $\scalebox{.8}{\bVee}\cdot \scalebox{.8}{\bVee}$ and 
$\scalebox{.8}{\bVee}^{\scalebox{.7}{\bVee}}$\,, have the same shape (the balanced, 4-leaf tree (figure \ref{fig:4leaf9j})), but of course with rearranged leaf labels according to $\ulj\cdot \ulj' = (j_1,j_2,j_1',j_2')$, 
$\ulj\ast\ulj' = (j_1,j_1',j_2,j_2')$. The recoupling coefficients between totally coupled angular momentum states according to the respective schemes are thus dependent on 
nine quantities: four leaf labels, two pairs of internal labels and the total angular momentum.

In section \ref{subsec:ExplicitNeq2products}, in the $N=2$ case, rather than a direct calculation of the quasicharacter product 
$\chibig_{j_1 j_2;k}(u,v) \chibig_{j'_1 j'_2;k'}(u,v)$\,, it was shown that 
the tensor product of the associated projectors
$\bP_{j_1j_2}^k \otimes\bP_{j'_1j'_2}^{k'} $ can be manipulated in such a way that the
quasicharacter product expansion follows immediately by taking an overall trace
See (\ref{eq:PtwiseNeq2Full}

As mentioned above, for specific evaluations it is convenient to refer
all products to a fixed coupling scheme. Consider for definiteness the specific choice 
based on the totally unbalanced or so-called ``caterpillar'' tree
$T^{\texttt{cat}}$; that is, with $N\!-\!2$ 
internal angular momenta assigned sequentially, viz.
$ j_{(12)}$\,, $j_{((12)3)}$\,, $\cdots$ $j_{(\cdots ((12)3)\cdots N\!-\!1})$\,, (labelled $k_{1i}\,,k_{2i}$\, below
 with $j := j_{(\cdots ((12)3)\cdots N)}$\,). This is of course 
identical to the standard Clebsch-Gordan $\ul j, \ul l$ scheme of equation (\ref{G-D-Hjl}) with, in this case,
$k_1=l_2\,, k_2=l_3\,, \cdots$\, , and $ k_{N\!-\!2}=l_{N\!-\!1}\,$ . The statement of 
lemma \ref{L-symbols}, in the composite tree labelling notation, becomes 
(c.f. \cite{Fuchs2018costratificationMR}, \cite{jarvis2021quasicharactersMR})
\begin{lem}[Pointwise product and $9j$ factorization] 
\label{lem:PtWise9jFactors}
The multiplicative structure constants for products of standard caterpillar tree quasicharacters are given (for $N \ge 3$\,) by
\begin{align*}
\label{eq:PtwiseNgen2Full}
&\,\chibig^{\texttt{cat},j_1}_{\ulj,\ulk_1\ulk'_1}(\ulu) \cdot 
\chibig^{\texttt{cat},j_2}_{\ulj,\ulk_2,\ulk'_2}(\ulu)= \sum_{j, \ulJ\ulK ,\ulK' }\chibig^{\texttt{cat},J}_{\ulJ, \ulK,\ulK'}(\ulu) \cdot \\
&\, 
\left[\left(\!\begin{array}{ccc} j_{11}& j_{21}&J_1 \\j_{12}& j_{22}&J_2\\ k_{11}& k_{21}& K_1\end{array}\!\right) \!\cdot\!\prod_{i=1}^{N\!-\!3}
 \left(\!\begin{array}{ccc}k_{1i}& k_{2i}& K_i \\
j_{1i\!+\!2}& j_{2i\!+\!2}& J_{i\!+\!2}\\ j_{1i\!+\!1}& j_{2i\!+\!1}& K_{i\!+\!1} \end{array}\!\right)
\!\cdot\! \left(\!\begin{array}{ccc}k_{1N\!-\!2}& k_{2N\!-\!2}& K_{N\!-\!2} \\
j_{1N\!-\!1}& j_{2N\!-\!2}& J_{N\!-\!1} \\ j_1& j_2 & J \end{array}\!\right)\right]
\!\cdot\! \left[\left(\!\begin{array}{c} \cdots \\ \cdots \\ \cdots \end{array}\!\right)\right]
\,,
\end{align*}
where the symbol $\left[(\!\begin{array}{c} \cdots \end{array}\!)\right]$ repeats the product
of $9j$ symbols (see Lemma \ref{L-symbols}), with $\ul k_1, \ul k_2, \ul K$ replaced by
$\ul k'_1, \ul k'_2, \ul K'$\,. For $N=2$\,, each product collapses to a single term yielding
the coefficient (as in lemma \ref{lem:PtWiseNeq2})\footnote{Identifiable using permutation symmetry on the $9j$ symbol rows and columns \,.})
\[ 
\left(\!\begin{array}{ccc} j_{11}& j_{21}&J_1 \\j_{12}& j_{22}&J_2\\ j_1& j_2 & J\end{array}\!\right)^{\!\!2}\,.
\]
\end{lem}
\noindent
\mbox{}\hfill $\Box$
\end{appendix}


\begin{table}
\label{tab:Neq2exx}
\begin{center}
\mbox{}\hskip -6ex
\mbox{}\vskip -10ex
\begin{tabular}{|c|l|}
\hline
\hline
& \\
$\left.{\strut}\right.^j_{(j_1j_2)}$ & 
$\frac{1}{\sqrt{[j_1][j_2]}}
\chibig^j_{(j_1j_2)} (u,v)$ \\&\\ 
\hline
\hline
&\\
$\left.{\strut}\right.{(\fh0)}^{\fh}$ &$Tr(u)$ \\
& \\
\hline
\hline
& \\
$\left.{\strut}\right._{(\fh\fh)}^{1}$ & ${\frac 12}\big(Tr(uv)+Tr(u)Tr(v) \big)$ \\
& \\
\hline
& \\
$\left.{\strut}\right._{(\fh\fh)}^{0}$ & ${\frac 12}\big(-Tr(uv)+Tr(u)Tr(v) \big)$ \\
& \\
\hline
& \\
$\left.{\strut}\right._{(10)}^{1}$ & $-1+Tr(u)^2$ \\
& \\
\hline \hline
& \\
$\left.{\strut}\right._{(\fh1)}^{\fh}$ & ${\frac 13}\big(-Tr(u)-2Tr(v)Tr(uv) +2Tr(u)Tr(v)^2\big)$ \\
& \\
\hline
& \\
$\left.{\strut}\right._{(\fh1)}^{\fth}$ & ${\frac 13}\big(-2Tr(u)+ 2Tr(v)Tr(uv) +Tr(u)Tr(v)^2 \big)$ \\
& \\
\hline
& \\
$\left.{\strut}\right._{(\fth0)}^{\fh}$ &$-2Tr(u)+ Tr(u)^3$ \\
& \\
\hline
\hline
& \\
$\left.{\strut}\right._{(11)}^{0}$ & ${\frac 13}\big(Tr(uv)^2 -2Tr(u)Tr(v)Tr(uv)+Tr(u)^2Tr(v)^2 -1\big)$ \\
& \\
\hline
& \\
$\left.{\strut}\right._{(11)}^{1}$ & ${\frac 12}\big(-Tr(u)^2-Tr(v)^2-Tr(uv)^2+Tr(u)^2Tr(v)^2 +2 \big)$ \\
& \\
\hline
& \\
$\left.{\strut}\right._{(11)}^{2}$ &$ {\frac 13} - {\frac 12}Tr(u)^2 - {\frac 12}Tr(v)^2 
+ {\frac 16}Tr(uv)^2 + {\frac 23}Tr(u)Tr(v)Tr(uv) + {\frac 16}Tr(u)^2Tr(v)^2\,$ \\
& \\
\hline
& \\
$\left.{\strut}\right._{(\fth\fh)}^{2}$ & $-  {\frac{1}{4}}Tr(uv)- {\frac{1}{4}}Tr(u)Tr(v) + {\frac{3}{4}}Tr(u)^2Tr(uv)- {\frac{3}{8}}Tr(u)^2Tr(v)+ {\frac{1}{16}}Tr(u)^3Tr(v)$ \\
& \\
\hline
& \\
$\left.{\strut}\right._{(\fth\fh)}^{1}$ & $+  {\frac{1}{4}}Tr(uv)+ {\frac{7}{4}}Tr(u)Tr(v) - {\frac{3}{4}}Tr(u)^2Tr(uv)+ {\frac{3}{8}}Tr(u)^2Tr(v)+ {\frac{15}{16}}Tr(u)^3Tr(v)$ \\
& \\
\hline
& \\
$\left.{\strut}\right._{(20)}^{2}$ & $1-3Tr(u)^2 + Tr(u)^4$ \\
& \\
\hline
\hline
\end{tabular}
\end{center}
\caption{\protect \small Tabulation of some $N=2$ quasicharacters $\chibig_{(j_1 j_2)}^{j} \big(u,v\big)$ up to degree 4. }
\end{table}

\begin{table}[tbp]
\mbox{}\\[-1cm]
\begin{center}
\begin{tabular}{|c|l|}
\hline
\hline
&\\
$\left.{\strut}\right.^j_{(j_1j_2j_3)l_2l_2'}$ & 
$\frac{1}{\sqrt{[j_1][j_2][j_3]}}
\chibig^j_{(j_1j_2j_3)l_2l_2'} (u,v,w)$ \\&\\ 
\hline
\hline
& \\
${\strut}_{(\fh\fh\fh)11}^{\fth}$ &
$\tfrac 16 \Big(
\Tr ( uwv  ) + \Tr (  vwu ) + \Tr ( uv  )\Tr ( w   ) +
\Tr ( uw  )\Tr ( v   ) + \Tr ( vw  )\Tr ( u   ) +\Tr (  u )\Tr (  v )\Tr (  w )\Big)$ \\
& \\
\hline
& \\
${\strut}_{(\fh\fh\fh)11}^{\fh}$ &$ \tfrac 16\Big(
-\Tr ( uwv  ) - \Tr (  vwu ) 
-
\Tr ( uw  )\Tr ( v   ) + 2\Tr ( uv  )\Tr ( w  )- \Tr ( vw  )\Tr ( u   )+ 
2\Tr (  u )\Tr (  v )\Tr (  w )\Big)$ \\
& \\
\hline
& \\
${\strut}_{(\fh\fh\fh)00}^{\fh}$ & 
$\tfrac 12\big(\Tr ( u )\Tr ( v ) \Tr ( w )- \Tr ( uv )
\Tr ( w )\big)$\\
&\\
\hline \hline
&\\
$\left.{\strut}\right.^j_{(j_1j_2j_3)l_2l_3,l_2'l_3'}$ & 
$\frac{1}{\sqrt{[j_1][j_2][j_3][j_4]}}
\chibig^j_{(j_1j_2j_3j_4)l_2l_3,l_2'l_3'} (r,s,t,u)$ \\&\\ 
\hline
\hline
& \\
$\left.{\strut}\right._{(\fh\fh\fh\fh)1\tfrac32,1\tfrac32}^{2}$ & 
\parbox{14cm}{$\frac{1}{24}\left\{\big(\frac 23 \Tr( rstu +rsut+ rust) + \frac 89
\Tr( rtsu+ rtus + ruts)
\big)\right.$ \\
$ + 
\big(\Tr( s )\Tr( rtu+rut) +\Tr( r )\Tr( stu+sut) 
\Tr( t )\Tr( rsu+rus)  +\frac 23 \Tr( u )\Tr( rst+rts) \big)$\\
$+\big(\Tr( rs ) \Tr( t) \Tr( u )
+\frac{13}{9}(\Tr( ru ) \Tr( s) \Tr( t )+\Tr( su ) \Tr( r) \Tr( t )) + 
\frac 49 \Tr( ut ) \Tr( r) \Tr( s )\big)$\mbox{}\hfill \\
$ \left.+ 2\big(\Tr( rs ) \Tr( ut) + \Tr( ru ) \Tr( st) +
\Tr( rt ) \Tr( su)\big)
+\frac{13}{9}\Tr( r ) \Tr( s) \Tr( t ) \Tr( u )\right\}\,.$}\\
& \\
\hline \hline
\end{tabular}
\end{center}
\caption{\protect \small Tabulation of some $N=3$ and $N=4$  quasicharacters.}
\label{tab:Neq3exx}
\end{table}
\mbox{}\\

\begin{table}[tbp]
\begin{center}\[
\begin{array}{|r | c c c c c c c c|}
\hline
\hline
&&&&&&&&\\
\chibig_{\scb{.85}{$\oh 0$}}^{\scb{.85}{$\oh$}}\cdot \chi_{\scb{.85}{$j \oh$}}^{\scb{.85}{$j\!+\!\oh$}} & 
1\cdot \chi_{\scb{.85}{$j\sp\oh \oh$}}^{\scb{.85}{$ j\sp 1$}} &
 &\displaystyle{\frac{1}{(2j\sp 1)^2}}\cdot \chi_{\scb{.85}{$j\sp\oh \oh$}}^{\scb{.85}{$ j$}}& &
\displaystyle{\frac{2j(2j\sp 2)}{(2j\sp 1)^2}}\cdot \chi_{\scb{.85}{$j\sm\oh \oh$}}^{\scb{.85}{ $j$}}& &0\cdot \chi_{\scb{.85}{$j\sm\oh \oh$}}^{\scb{.85}{ $j\sm 1$}}&  \\  
&&&&&&&&\\
\hline
&&&&&&&&\\
\scb{.85}{$\dtnj$} &\displaystyle{\frac{1}{2(2j\sp 2)}}&& 
\displaystyle{\frac{1}{2(2j\sp 1)(2j\sp 2)}}&&
\displaystyle{\frac{1}{2(2j\sp 1)}}&&0&\\
&&&&&&&&\\
\hline
&&&&&&&&\\
& \scb{.8}{$\nj{\oh}{0}{\oh}{j}{\oh}{j\sp\oh}{j\sp \oh }{\oh }{j \sp 1}$}&&
\scb{.8}{$\nj{\oh}{0}{\oh}{j}{\oh}{j\sp\oh}{j\sp \oh }{\oh }{j}$}&&
\scb{.8}{$\nj{\oh}{0}{\oh}{j}{\oh}{j\sm\oh}{j\sm \oh }{\oh }{j}$}&&
\scb{.8}{$\nj{\oh}{0}{\oh}{j}{\oh}{j\sm\oh}{j\sm \oh }{\oh }{j\sm 1}$}
&\\
&&&&&&&&\\
\hline
\hline
\end{array}
\]
\end{center}
\caption{ $9j$ symbols derived from the expansion of $N=2$ quasicharacter products.}
%
\label{tab:9jcalcs}
\end{table}
\begin{table}[tbp]
\begin{center}
\[
\begin{array}{|r | c c c c c c |}
\hline
\hline
&&&&&&\\
\chi_{\scb{.85}{$0 \oh$}}^{\scb{.85}{$\oh$}}\cdot \chi_{\scb{.85}{$j \oh$}}^{\scb{.85}{$j\!+\!\oh$}} & 
1\cdot \chi_{\scb{.85}{$j  1$}}^{\scb{.85}{ $j\sp 1$}} &
 &\displaystyle{\frac{j}{(2j\sp 1)}}\cdot \chi_{\scb{.85}{$j  1$}}^{\scb{.85}{$ j\sm 1$}}& &
\displaystyle{\frac{(j\sp 1)}{(2j\sp 1)}}\cdot \chi_{\scb{.85}{$j 0$}}^{\scb{.85}{ $j$}} & \\  
&&&&&&\\
\hline
&&&&&&\\
\scb{.85}{$\dtnj$} &\displaystyle{\frac{1}{2\sqrt{3(j\sp 1)(2j\sp 1)}}}&& 
\displaystyle{\frac{1}{2(2j\sp 1)}\sqrt{\frac{j}{3(j\sp 1)} } }&&
\displaystyle{\frac{1}{2(2j\sp 1)}\sqrt{\frac{1}{2j\sp 1}}}&\\
&&&&&&\\
\hline
&&&&&&\\
& \scb{.8}{$\nj{0}{\oh}{\oh}{j}{\oh}{j\sp\oh}{j  }{1 }{j \sp 1}$}&&
\scb{.8}{$\nj{0}{\oh}{\oh}{j}{\oh}{j\sp\oh}{j }{1 }{j}$}&&
\scb{.8}{$\nj{0}{\oh}{\oh}{j}{\oh}{j\sp\oh}{j }{0 }{j}$}&\\
&&&&&&\\
\hline
\hline
\end{array}
\]
\end{center}
\caption{ $9j$ symbols derived from the expansion of $N=2$ quasicharacter products (ctd.).}
\label{tab:9jcalcsctd}
\end{table}

\begin{table}[tbp]
\mbox{}\hskip 3cm
\begin{tabular}{|c|l|l|c|}
\hline \hline
&&\\
$n$ & \parbox{2cm}{Monomial\\ 
\& number} & Total \\
&&\\
\hline \hline 
&&\\
2 &$\begin{array}{llr}
(1^2)&1\\(2)&1 \end{array}$&2\\
&&\\
\hline
&&\\
3 &$\begin{array}{llr}
(1^3)&1\\(21)&3 \\3&1 \end{array}$& 5\\
&&\\
\hline
&&\\
4 &$\begin{array}{llr}
(1^4)&1\\(21^4)&6 \\(2^2)&3 \\(31) & 4\end{array}$ & 14\\
&&\\
\hline \hline
\end{tabular}
\hskip 3ex
\begin{tabular}{|c|l|l|c|}
\hline \hline
&&\\
$n$ & \parbox{2cm}{Monomial\\ 
\& number} & Total \\
&&\\
\hline \hline 
&&\\
5 &$\begin{array}{llr}
(1^5)&1\\(21^3)&10\\(31^2)&10\\(2^21)&15\\(32)& 6^* \end{array}$
& 42\\
&&\\
\hline&&\\
6 &$\begin{array}{llr}
(1^6)&1\\(21^4)&15\\(31^3)&20\\(2^21^2)&45\\(2^3)&15\\(321)& 36^* \\(33)&0^* \end{array}$
& 132\\
&&\\
\hline \hline
\end{tabular}
\caption{\protect \small 
The Catalan number $C(n)$ of monomials in linear, quadratic and cubic traces at degree $n=3p\!+\!2q\!+\!r$\,. 
}
\label{tab:MonomialCount}
\end{table}

\begin{table}
\begin{tabular}{|c|ll|}
\hline \hline
&&\\[-.3cm]
$H_{\underline{j}}$ & irreducible $\times^N{\SU(2)}$ module ($=H_{j_1}\otimes H_{j_2}\otimes \cdots \otimes H_{j_N}$)&\\
&&\\[-.3cm]
\hline
&&\\[-.3cm]
$d_j=2j+1 \equiv{[}j{]}$ & dimension of irreducible ${\SU(2)}$ module (spin $j$\,,
Dynkin label $2j$)&\\
&&\\[-.3cm]
\hline
&&\\[-.3cm]
$H_{j, \ul j}$ & isocopy space, $= H_j^{(1)} \! + \! H_j^{(2)} \! + \! \cdots \! + \! H_j^{(r)}$\,, $H_j^{(a)} \cong H_j\,, a=1,2,\cdots, r$ &\\
&&\\[-.3cm]
\hline
&&\\[-.3cm]
$m_{\underline{j}}(j) $ & multiplicity of
$H_j$ within $H_{\ul j, j}$ $(=r)$ &\\
&&\\[-.3cm]
\hline
&&\\[-.3cm]
$\ul j, \ul l, j $ & spin labels (standard coupling; $l_1\equiv j_1$\,,$l_N \simeq j$ )&\\
&&\\[-.3cm]
\hline
&&\\[-.3cm]
$T, \ul j, \ul k, j$& spin labels (tree, leaf edges,  internal edges, root edge) &\\ 
&&\\[-.3cm]
\hline
&&\\[-.3cm]
$\mid(j_1 j_2) j_{12},m\rangle$ & two spin coupled state &\\
&&\\[-.3cm]
\hline
&&\\[-.3cm]
$\mid {\ul j}, {\ul l}, j,m\rangle$& multi-spin coupled state (standard coupling) &\\
&&\\[-.3cm]
\hline
&&\\[-.3cm]
$\mid T({\ulj}, {\ulk}), j,m\rangle$& multi-spin coupled state (general $T$) &\\
&&\\[-.3cm]
\hline
&&\\[-.3cm]
$\BF{\ul j}{j}{\ul l,\ul l'}(\ul u)$ & $\SU(2)$ quasicharacter (standard coupling)&\\
&&\\[-.3cm]
\hline
&&\\[-.3cm]
${\chibig}^{T,j}_{(\ulj),\ulk,\ulk'}({\ulu})$ & $\SU(2)$ quasicharacter (general $T$)&\\
&&\\[-.3cm]
\hline
&&\\[-.3cm]
${\chibig}^j_{k l}(u,v)$ & $N=2$ quasicharacter &\\
&&\\[-.3cm]
\hline &&\\[-.3cm]
\,${\chibig}^{j}_{(j_1j_2j_3)k, k'}(u,v,w)$ & $N=3$ quasicharacter&\\
&&\\[-.3cm]
\hline
&&\\[-.3cm]
$C^{j_1,j_2, j_{12}}_{m_1,m_2,m_{12}}$& Clebsch-Gordan coefficient $(=\langle j_1m_1,j_2m_2\mid (j_1j_2) j_{12},m\rangle$\,) & \\
&&\\[-.3cm]
\hline
&&\\[-.3cm]
$C(T)^{\ulj}_{{\ulm}; {\ulk}, j, m }$&  Clebsch-Gordan coefficient $(=\langle \ul j, \ul m |T(\ulj ,\ul k) jm \rangle) $&\\
&&\\[-.3cm]
\hline
&&\\[-.3cm]
$R\big(T'\mid T\big)^{\ulj,j}_{\ulk', \ulk}$ &
Racah recoupling coefficient ($ =\langle T'(\ulj, \ulk'); j m | T(\ulj ,\ulk); jm \rangle$)&\\
&&\\[-.3cm]
\hline
&&\\[-.3cm]
$\left.\left\{\begin{array}{ccc} j_1 & j_2 & k\\j'_1 & j'_2 & k' \\
J_1 & J_2 & K \end{array}\right\}\right.   $ &  Racah $9j$ symbol  &  \\
&&\\[-.3cm]
\hline
&&\\[-.3cm]
${\mathbb P}^j_{j_1,j_2}$& two spin projection operator &\\
&&\\[-.3cm]
\hline
&&\\[-.3cm]
${\mathbb P}(T)^j_{\ulj,\ulk}$& multi-spin projection operator &\\
&&\\[-.3cm]
\hline
&&\\[-.3cm]
${\mathbf J}=D_{j}(\textstyle {\frac 12}\boldsig)$\,& spin-$j$ generator &\\ 
&&\\[-.3cm]
\hline
&&\\[-.3cm]
${\mathbb K}_{(123)}$ & permutation operator ($= {\mathbb K}_{(23)} \circ {\mathbb K}_{(12)}$)&\mbox{} \hskip 2cm \\
&&\\[-.3cm]
\hline
&&\\[-.3cm]
${\mathbb S}^j$ & $j$-fold symmetrization operator (for $\pi_j(u)$)&\mbox{} \hskip 2cm \\
&&\\[-.3cm]
\hline
&&\\[-.3cm]
${\mathbb A}^2, {\mathbb M}^{j+1}$ & antisymmetrization, 
mixed symmetrization operator (for $\pi_{(j\!-\!1)}(u)$)&\mbox{} \hskip 2cm \\
&&\\[-.3cm]
\hline
\hline
\end{tabular}
\caption{\protect \small  Index of notation.}
\label{tab:NotationIndex}
\end{table}

\end{document}